\documentclass[twocolumn]{aastex62}
\hypersetup{breaklinks}

\newcommand{\caii}{\ion{Ca}{2}}
\newcommand{\caiihk}{\ion{Ca}{2} H \& K}

\newcommand{\rphk}{\ensuremath{R'_{\rm HK}}}
\newcommand{\lrphk}{\ensuremath{\log{\rphk}}} 

\newcommand{\feh}{\ensuremath{[\mbox{Fe}/\mbox{H}]}}
\newcommand{\mgb}{\ion{Mg}{1}~b}
\newcommand{\teff}{\ensuremath{T_{\mbox{\scriptsize eff}}}}
\newcommand{\logg}{\ensuremath{\log g}}
\newcommand{\msun}{\ensuremath{\mbox{M}_{\odot}}}
\newcommand{\rsun}{\ensuremath{\mbox{R}_{\odot}}}
\newcommand{\mearth}{\ensuremath{\mbox{M}_{\earth}}}
\newcommand{\rearth}{\ensuremath{\mbox{R}_{\earth}}}
\newcommand{\mjup}{\ensuremath{\mbox{M}_{\rm Jup}}}

\newcommand{\prot}{\ensuremath{P_{\mbox{\scriptsize rot}}}}
\newcommand{\vsini}{\ensuremath{v \sin i}}
\newcommand{\kms}{\ensuremath{\mbox{km s}^{-1}}}
\newcommand{\mps}{\ensuremath{\mbox{m s}^{-1}}}

\newcommand{\soren}{S\o ren Meibom}
\newcommand{\degree}{\ensuremath{^\circ}}
\newcommand{\mas}{\ensuremath{\mbox{mas yr}^{-1}}}

\newcommand{\Rnom}{\hbox{$(\mathcal{R})^{\rm N}_{\odot}$}}
\newcommand{\REnom}{\hbox{$(\mathcal{R})_{\earth}^{\rm N}$}}

\newcommand\blender{{\tt BLENDER}}

\newcommand{\thisepic}{EPIC~219800881}
\newcommand{\thisstar}{K2-231}
\newcommand{\thisplanet}{K2-231\,b}

\newcommand{\SpecVsini}{2.0}
\newcommand{\eSpecVsini}{0.5}

\newcommand{\StarMass}{1.01}
\newcommand{\eStarMass}{0.03}

\newcommand{\StarRadius}{0.95}
\newcommand{\eStarRadius}{0.03}

\newcommand{\StarTeff}{5695}
\newcommand{\eStarTeff}{50}

\newcommand{\StarLogg}{4.48}
\newcommand{\eStarLogg}{0.03}

\newcommand{\PerOrb}{13.841901}
\newcommand{\ePerOrb}{0.001352}

\newcommand{\RpRs}{0.0239}
\newcommand{\eRpRs}{$^{+0.0020}_{-0.0012}$}

\newcommand{\ScaledSemi}{27.0}
\newcommand{\eScaledSemi}{$^{+4.8}_{-4.1}$}

\newcommand{\Rplanet}{2.5}
\newcommand{\eRplanet}{0.2}

\newcommand{\Incl}{88.6}
\newcommand{\eIncl}{$^{+0.9}_{-0.6}$}

\newcommand{\Impact}{0.55}
\newcommand{\eImpact}{$^{+0.23}_{-0.37}$}

\newcommand{\Tdur}{2.94}
\newcommand{\eTdur}{$^{+2.02}_{-1.15}$}

\newcommand{\TransitTime}{57320.00164}
\newcommand{\eTransitTime}{$^{+0.00354}_{-0.00340}$}

\received{November 3, 2017}
\revised{February 1, 2018}
\accepted{March 4, 2018}
\submitjournal{AJ}

\shorttitle{A sub-Neptune exoplanet in Ruprecht 147}
\shortauthors{Curtis et al.}

\begin{document}

\title{K2-231\,\MakeLowercase{b}: A sub-Neptune exoplanet transiting a solar twin in Ruprecht 147}

\newcommand{\cehw}{Center for Exoplanets and Habitable Worlds, Department of Astronomy \& Astrophysics, 
    The Pennsylvania State University, \\ 
    525 Davey Laboratory, University Park, PA 16802, USA}
\newcommand{\columbia}{Department of Astronomy, Columbia University, 550 West 120th Street, New York, NY 10027, USA}
\newcommand{\texas}{Department of Astronomy, The University of Texas at Austin, Austin, TX 78712, USA}
\newcommand{\cfa}{Harvard--Smithsonian Center for Astrophysics, 60 Garden Street, Cambridge, MA 02138, USA}
\newcommand{\berkeley}{Astronomy Department, University of California, Berkeley, CA, USA}
\newcommand{\caltech}{Department of Astronomy, California Institute of Technology,
Pasadena, CA, USA}
\newcommand{\Hawaii}{Institute for Astronomy, University of Hawai`i, 2680 Woodlawn Drive, Honolulu, HI 96822, USA}
\newcommand{\seti}{SETI Institute, 189 Bernardo Avenue, Mountain View, CA 94043, USA}
\newcommand{\sydney}{Sydney Institute for Astronomy (SIfA), School of Physics, University of Sydney, NSW 2006, Australia}
\newcommand{\sac}{Stellar Astrophysics Centre, Department of Physics and Astronomy, Aarhus University, Ny Munkegade 120, DK-8000 Aarhus C, Denmark}
\newcommand{\cahill}{Cahill Center for Astrophysics, California Institute of Technology, Pasadena, CA, USA}
\newcommand{\ames}{NASA Ames Research Center, Moffett Field, CA 94035, USA}

\correspondingauthor{Jason Lee Curtis}
\email{jasoncurtis.astro@gmail.com}

\author[0000-0002-2792-134X]{Jason Lee Curtis}
\altaffiliation{NSF Astronomy and Astrophysics Postdoctoral Fellow}
\affiliation{\columbia}
\affiliation{\cehw}
\affiliation{\cfa}    
    
\author{Andrew Vanderburg}
\altaffiliation{NASA Sagan Fellow}
\affiliation{\cfa}
\affiliation{\texas}

\author{Guillermo Torres}
\affiliation{\cfa}

\author{Adam L. Kraus}
\affiliation{\texas}

\author{Daniel Huber}
\affiliation{\Hawaii}
\affiliation{\sydney}
\affiliation{\seti}
\affiliation{\sac}

\author{Andrew W.~Mann}
\altaffiliation{NASA Hubble Fellow}
\affiliation{\columbia}
\affiliation{\texas}

\author{Aaron C. Rizzuto}
\affiliation{\texas}

\author{Howard Isaacson}
\affiliation{\berkeley}

\author{Andrew W. Howard}
\affiliation{\caltech}

\author{Christopher E. Henze}
\affiliation{\ames}

\author{Benjamin J. Fulton}
\affiliation{\caltech}

\author{Jason T. Wright}
\affiliation{\cehw}


\begin{abstract}
We identify a sub-Neptune exoplanet ($R_p = 2.5 \pm 0.2$~\rearth) transiting a solar twin in the 
Ruprecht 147 star cluster (3 Gyr, 300 pc, [Fe/H] = +0.1 dex). 
The $\sim$81 day light-curve for EPIC 219800881 ($V = 12.71$) from \textit{K2} Campaign~7 shows 
six transits with a period of 13.84~days, a depth of $\sim$0.06\%, and a 
Based on our analysis of high-resolution MIKE spectra,  
broadband optical and NIR photometry, 
the cluster parallax and interstellar reddening, 
and isochrone models from PARSEC, Dartmouth, and MIST, 
we estimate the following properties for the host star: 
$M_\star =$ \StarMass\ $\pm$ \eStarMass~\msun,
$R_\star =$ \StarRadius\ $\pm$ \eStarRadius~\rsun, and
$\teff =$ \StarTeff\ $\pm$ \eStarTeff~K.
This star appears to be single 
based on our modeling of the photometry,
the low radial velocity (RV) variability measured over 
nearly ten years, 
and Keck/NIRC2 adaptive optics imaging and aperture-masking interferometry.
Applying a probabilistic mass--radius relation, 
we estimate that the mass of this planet is $M_p = 7 +5 -3$~\mearth , 
which would cause an RV semi-amplitude of $K = 2 \pm 1$~\mps\
that may be measurable with existing precise RV facilities. 
After statistically validating this planet with \blender, 
we now designate it K2-231\,b, 
making it the second substellar object to be discovered in Ruprecht 147 and the first planet;
it joins the small but growing ranks of 22 other planets and 3 candidates found in open clusters.
\end{abstract}




\section{Introduction} 
Transit and Doppler surveys have detected thousands of exoplanets,\footnote{As of 2017 June 9, 2950 were confirmed with 2338 additional \textit{Kepler} candidates; \url{http://exoplanets.org}}
and modeling their rate of occurrence shows that approximately 
one in three Sun-like stars hosts at least one planet with an orbital period under 29 days \citep{Fressin2013}.
Stars tend to form in clusters 
from the gravitational collapse and fragmentation of molecular clouds
\citep{LadaLada2003}, 
so it is natural to expect that stars still existing in clusters likewise 
host planets at a similar frequency. 
In fact, circumstellar disks have been observed in very young clusters and moving groups
\citep[2.5--30~Myr;][]{Haisch2001}.
However, some have speculated that stars forming in denser cluster environments 
(i.e., the kind that can remain gravitationally bound for billions of years) 
will be exposed to harsher conditions than stars formed in looser associations 
or that join the Galactic field relatively quickly after formation, 
and this will impact the frequency of planets formed and presently existing in star clusters. 
For example, stars in a rich and dense cluster might experience multiple supernovae 
during the planet-forming period (the lifetime of a 10~\msun\ star is $\sim$30~Myr), 
as well as intense FUV radiation from their massive star progenitors that can photoevaporate disks. 
Furthermore, stars in denser clusters ($\sim$0.3--30 FGK stars pc$^{-3}$)\footnote{Based on 528 single 
and binary members in M67 contained within 7.4~pc 
and 111 members within the central 1~pc \citep{Geller2015}.}
will also dynamically interact with other stars (and binary/multiple systems) 
at a higher frequency than more isolated stars in the field 
($\sim$0.06 stars pc$^{-3}$),\footnote{Based on the 259 systems within 10~pc 
tabulated by the REsearch Consortium On Nearby Stars 
\citep[RECONS;][]{Henry1997, Henry2006}; \url{http://www.recons.org/}}
which might tend to disrupt disks and/or eject planets from their host star systems. 

These concerns have been addressed theoretically and 
with observations 
\citep{Scally2001, Smith2001, Bonnell2001, Adams2006, Fregeau2006, Malmberg2007, Spurzem2009, Ovelar2012, Vincke2016, Kraus2016, Cai2017}, 
and all of these factors were considered by \citet{Adams2010} 
in evaluating the birth environment of the solar system, 
but progress in this field necessitates that we 
actually detect and characterize planets in star clusters 
and determine their frequency of occurrence. 

\subsection{Planets Discovered in Open Clusters}

Soon after the discovery of the first known exoplanet orbiting a Sun-like star \citep{Mayor1995}, 
\citet{Janes1996} suggested open clusters as ideal targets for photometric monitoring. 
Two decades later,  we still only know of a relatively small number 
of exoplanets existing in open clusters.
One observational challenge has been that the majority of nearby star clusters 
are young, and therefore their stars are rapidly rotating and magnetically active. 
Older clusters with inactive stars tend to be more distant, 
and their Sun-like stars are likewise too faint for most Doppler and ground-based transit facilities. 
The first planets discovered in open clusters with the Doppler technique 
were either massive Jupiters or potentially brown dwarfs:
\citet{LovisMayor2007} found two substellar objects in NGC~2423 and NGC~4349;\footnote{The 
substellar objects have minimum masses of 10.6 and 19.8~\mjup , respectively. 
\citet{Spiegel2011} calculated the deuterium-burning mass limits for brown dwarfs 
to be 11.4--14.4~\mjup , which supports a brown dwarf classification for the later object 
and places the former on the boundary between regimes.}
\citet{Sato2007} detected a companion to a giant star in the Hyades; 
\citet{Quinn2012} discovered two hot Jupiters in Praesepe 
\citep[known as the ``two b's in the Beehive,'' one of which also has a Jupiter-mass planet in a long-period, eccentric orbit;][]{Malavolta2016},
and 
\citet{Quinn2014} discovered another in the Hyades; 
and, finally, nontransiting hot Jupiters have been found in M67 
around three main-sequence stars, one Jupiter was detected around an evolved giant, 
and three other planet candidates were identified \citep{Brucalassi2014, Brucalassi2016, Brucalassi2017}.

NASA's \textit{Kepler} mission changed this by providing high-precision photometry 
for four clusters \citep{Meibom2011}. 
Two sub-Neptune-sized planets were discovered in the 1 Gyr NGC 6811 cluster, 
and \citet{Meibom2013} concluded that planets occur in that dense environment ($N = 377$ stars)
at roughly the same frequency as in the field. 
After \textit{Kepler} was repurposed as \textit{K2}, 
many more clusters were observed for $\sim$80 days each, 
and as a result, many new cluster planets have been identified.
Many of these are hosted by lower-mass stars that are intrinsically faint and 
difficult to reach with existing precision radial velocity (RV) facilities from Earth.
So far, results have been reported from \textit{K2} monitoring of the following clusters, 
listed in order of increasing age:
\citet{Gaidos2017} reported zero detections in the Pleiades \citep[see also][]{Mann2017}.
\citet{Mann2016} and \citet{David2016} independently discovered 
a Neptune-sized planet transiting a M4.5 dwarf in the Hyades; 
recently 
\citet{Mann2017Hyades} reported three Earth-to-Neptune-sized planets orbiting a mid-K dwarf 
in the Hyades (K2-136), while \citet{Ciardi2017} concurrently 
announced the Neptune-sized planet and that this K dwarf formed a binary with a late-M dwarf;
the system was later reported on by \citet{Livingston2017}.\footnote{As we are listing only 
validated exoplanets, 
we do not include polluted white dwarfs, like the one in the Hyades \citep{Zuckerman2013}.}
In Praesepe, 
\citet{Obermeier2016} announced K2-95~b, a Neptune-sized planet orbiting an M dwarf, 
which was later studied by \citet{Libralato2016}, \citet{Mann2017}, and \citet{Pepper2017}; 
adding the planets found by \citet{Pope2016}, \citet{Barros2016}, \citet{Libralato2016}, and \citet{Mann2017}, 
there are six confirmed planets (including K2-100 through K2-104) and one candidate 
that were validated by \citet{Mann2017}.
Finally, \citet{Nardiello2016} reported three planetary candidates in the M67 field,
although all appear to be nonmembers.

Table~\ref{t:cp} lists the 23 planets and 3 candidates that have been discovered in clusters so far, including \thisplanet.\footnote{
In this list, we have neglected exoplanets found in young associations 
and moving groups like 
Upper Sco \citep{Mann2016UpSco, David2016UpSco}, 
Taurus--Auriga 
\citep{Donati2016, Donati2017, Yu2017}, 
and 
Cas--Tau \citep{David2018}.}
Of these, 
15 transit their host stars, and all but six of the hosts are fainter than $V>13$, 
which makes precise RV follow-up prohibitively expensive. 
These hosts are all relatively young ($\sim$650~Myr) 
and magnetically active
and thus might still present a challenge to existing Doppler facilities and techniques.
Such RV observations are required to measure masses and determine the densities of these planets. 

\subsection{The \textit{K2} Survey of Ruprecht 147}

Ruprecht 147 was also observed by \textit{K2} during Campaign 7.\footnote{J.~Curtis 
successfully petitioned to reposition the Campaign~7 field 
in order to accommodate R147, 
which would have been largely missed in the originally proposed pointing.}
\citet{Curtis2013} demonstrated that R147 is the oldest nearby star cluster, 
with an age of 3~Gyr at a distance of 300~pc 
\citep[see also the Ph.D. dissertation of][]{Curtis2016PhD}.
According to \citet{Howell2014}, planets only a few times larger in size than Earth
would be detectable around dwarfs at least as bright as $K_p < 16$, 
which approximately corresponds to an M0 dwarf with $M = 0.6$~\msun\ in R147 . 
Soon after the public release of the Campaign 7 light-curves, 
we discovered a substellar object transiting a solar twin in Ruprecht 147 
(EPIC~219388192; CWW~89A from \citeauthor{Curtis2013} \citeyear{Curtis2013}), 
which we determined was a warm brown dwarf in an eccentric $\sim$5~day orbit, 
and we announced our discovery at 
the 19th Cambridge Workshop on Cool Stars, Stellar Systems, and the Sun (``Cool Stars 19'') 
in Uppsala, Sweden \citep{BDposter2016}.\footnote{\url{https://doi.org/10.5281/zenodo.58758}}
\citet{Nowak2017} independently discovered and characterized this system.

Now we report the identification of an object transiting a different solar twin in R147 \citep[CWW~93 from][]{Curtis2013}, 
which we show is a sub-Neptune exoplanet.
 We made this discovery while reviewing and comparing light-curves from various groups
for our stellar rotation program
\added{(we are measuring rotation periods for R147's FGKM dwarfs to validate and calibrate gyrochronology at 3 Gyr)}, 
and noticed a repeating shallow transit pattern spaced at $\sim$14 days in the 
\texttt{EVEREST} light-curve for \thisepic\
\citep{EVEREST1, EVEREST2}.\footnote{\url{https://archive.stsci.edu/prepds/everest/}}
%

In this paper, 
we describe our production of a light-curve, which we 
model to derive the properties of the exoplanet (Section \ref{s:k2}). 
We also characterize the host star and check for stellar binary companionship (Section \ref{s:star}), 
and test false-positive scenarios in order to statistically validate the exoplanet (Section \ref{s:false}). 

\thisstar\ was also targeted by the following programs: 
``Statistics of Variability in Main-Sequence Stars of Kepler 2 Fields 6 and 7''	(PI: Guzik; GO 7016),
``The Masses and Prevalence of Small Planets with K2 -- Cycle 2'' (PI: Howard; GO 7030), and
``K2 follow-up of the nearby, old open cluster Ruprecht 147'' (PI: Nascimbeni; GO 7056).

\section{\textit{K2} Light-curve Analysis} \label{s:k2}

\begin{figure*}\begin{center}
\includegraphics[width=6in, trim=0 0 0 1cm]{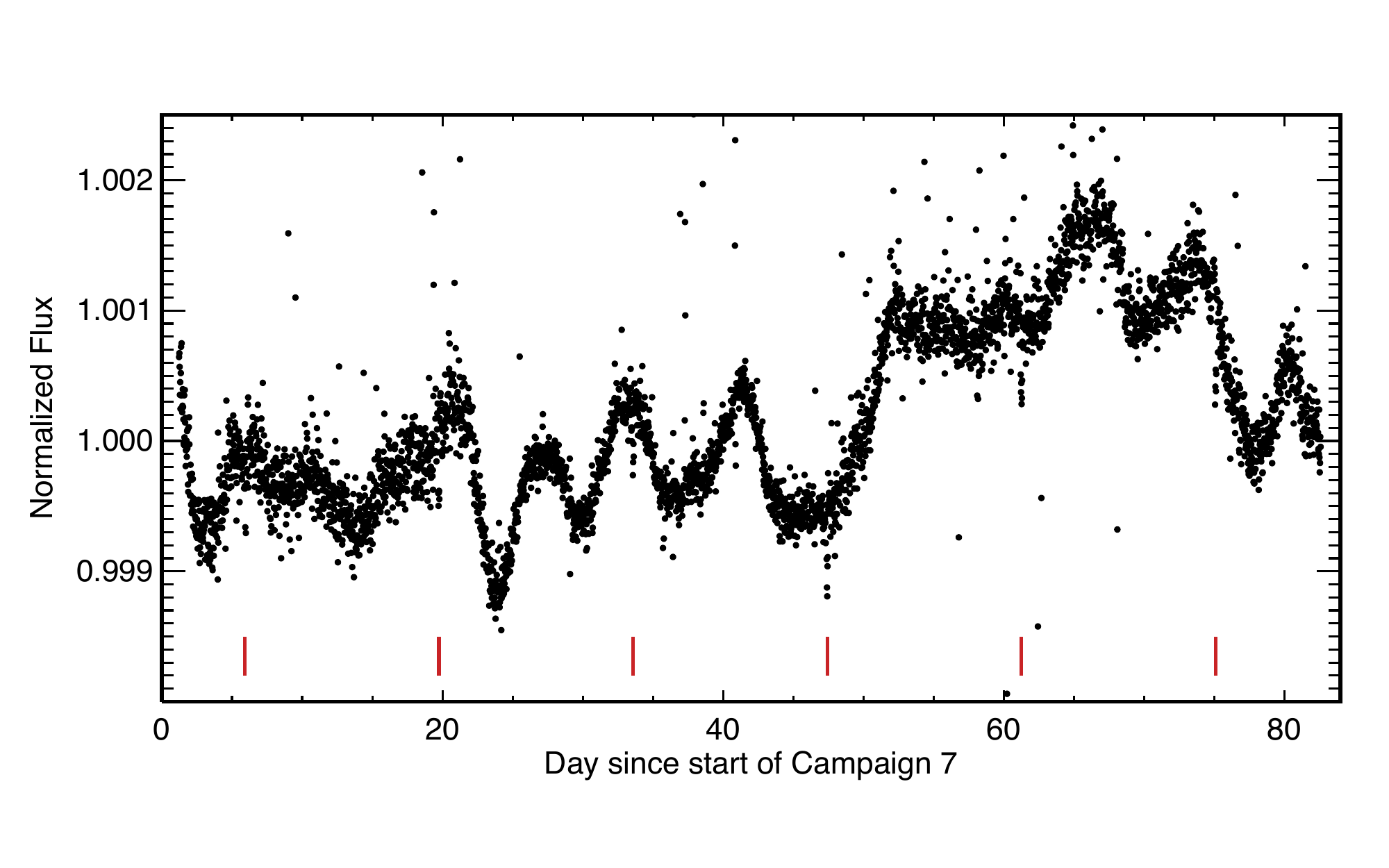}
\includegraphics[width=6in, trim=0 0 0 3cm]{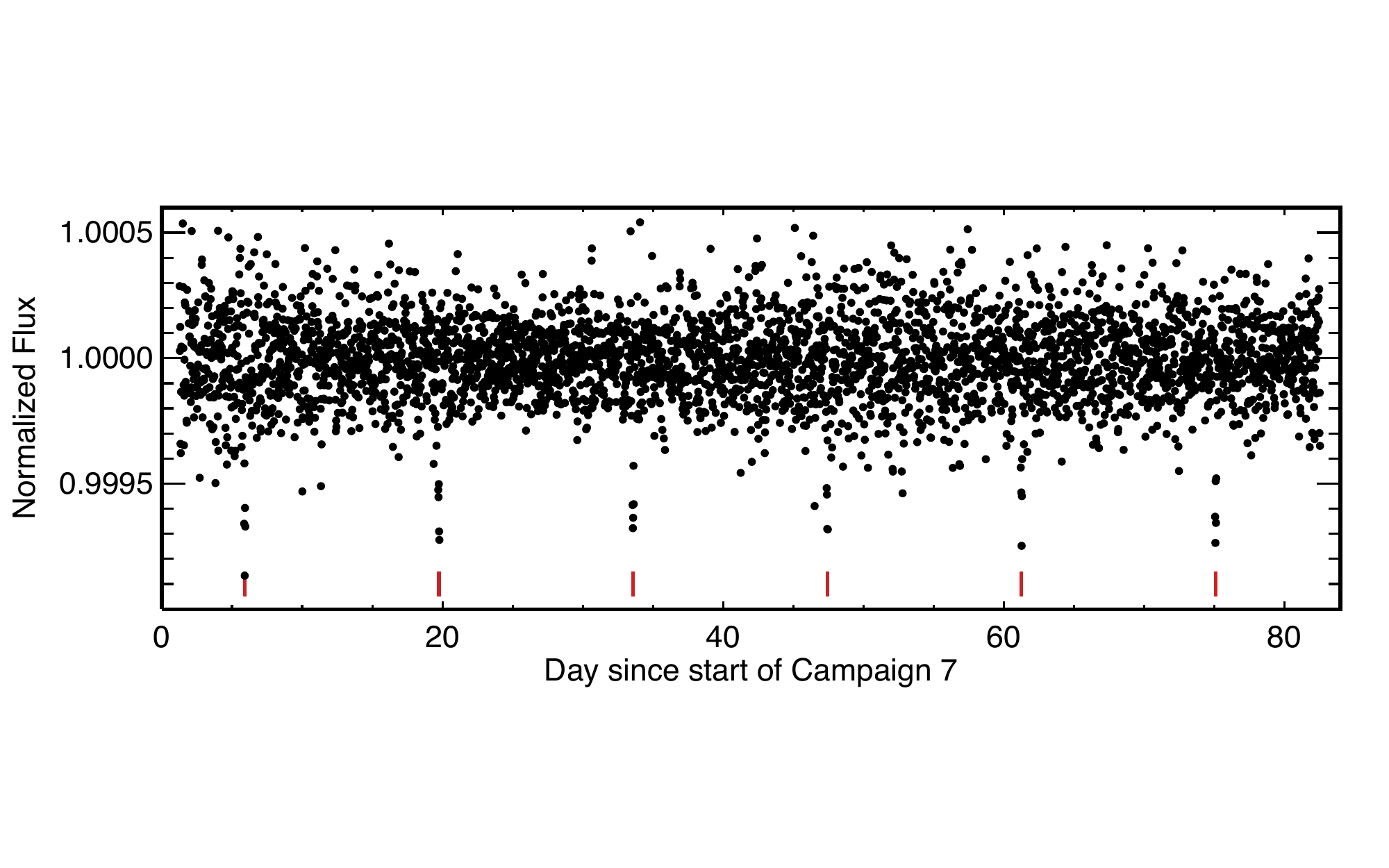}
\includegraphics[width=4in, trim=0 0 0cm 2cm]{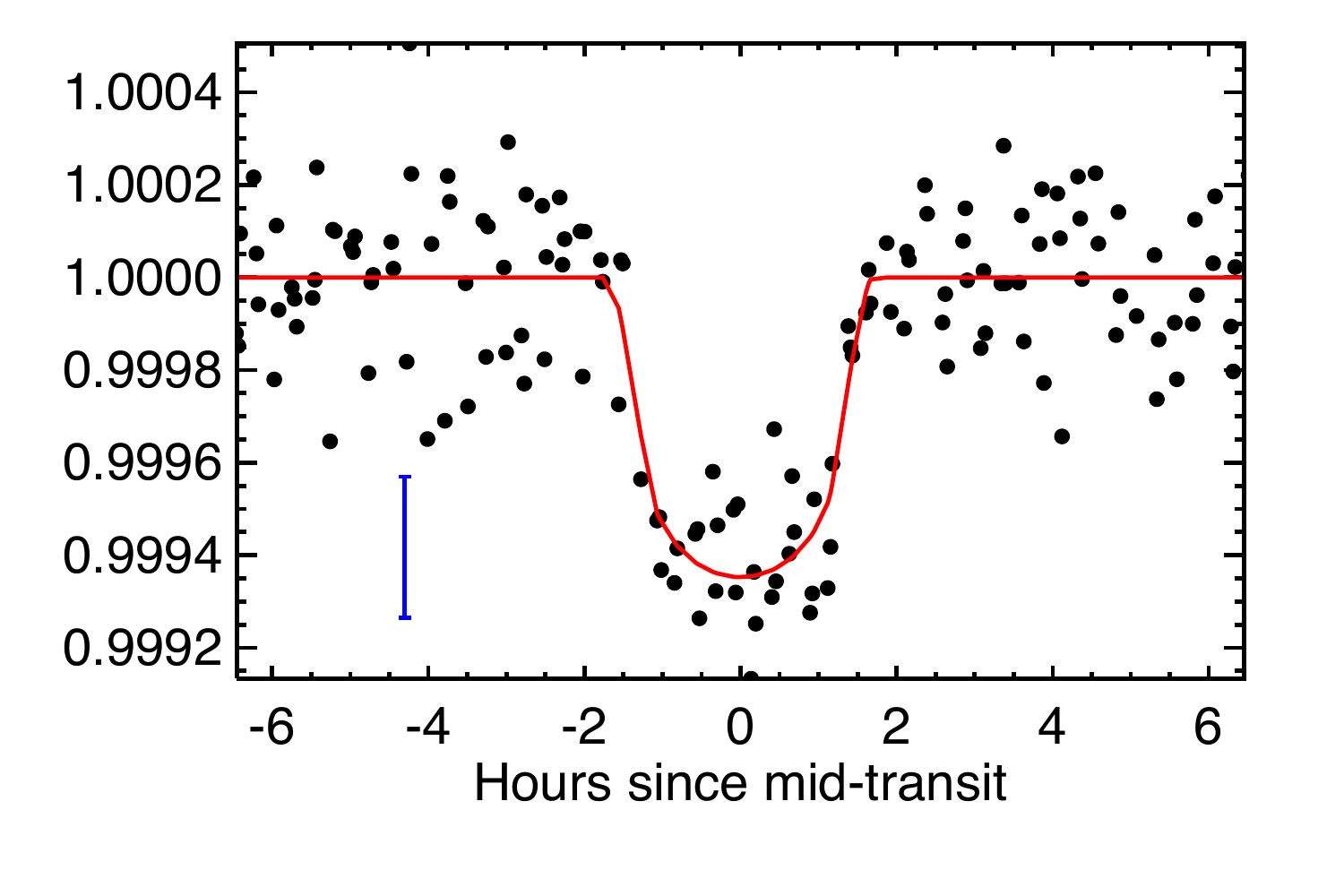}
\vspace{-0cm}
	\caption{\textit{K2} light-curves for \thisepic: 
	(Top) The \texttt{EVEREST} light-curve used to visually identify the transiting object, 
	with the transits marked as short red vertical lines at the bottom of the figure.
	(Middle) Our refined and detrended light-curve, extracted with a $9''$ circular moving aperture while 
	simultaneously fitting the pointing systematics, activity signal, and transits following \citet{vanderburg2016}, with the transits similarly marked.
	(Bottom) Our detrended light-curve, phase-folded according to the 13.842~day period, 
	along with the model for highest-likelihood solution from 
	the ZEIT transit-fit procedure (see Table~\ref{t:prop}), sampled at the times of observation according to the 30 minute integration cadence (red).
	\added{Our calibrated light-curve and the detrended version 
	are both available in the online journal and the arXiv posting.}
	\label{f:lc}}
\end{center}\end{figure*}

The top panel of Figure \ref{f:lc} shows the \texttt{EVEREST} light-curve for \thisepic\ 
that caught our attention.
We then downloaded the calibrated pixel-level data from 
the Barbara A. Mikulski Archive for Space Telescopes 
(MAST),\footnote{\url{https://archive.stsci.edu/k2/}}
extracted a light-curve, 
and corrected for \textit{K2} systematic effects following \citet{vj14}. 
We confirmed the transits detected by eye with a Box-fitting Least-Squares (BLS) periodogram search \citep{kovacs}.\footnote{We made the original period measurement with the Periodogram Service available at \url{https://exoplanetarchive.ipac.caltech.edu}}
The BLS periodogram identified a strong signal at a 13.844~day period with a transit depth of approximately 0.06\%. 
We then refined the light-curve by simultaneously fitting the \textit{K2} pointing systematics,
a low-frequency stellar activity signal 
(modeled with a basis spline with breakpoints spaced every 0.75 days),
and transits (using \citeauthor{Mandel2002}~\citeyear{Mandel2002} models), 
as described in Section 4 of \citet{vanderburg2016}. 
Deviating from our standard procedure of using stationary apertures, we opted to use a smaller, moving circular aperture
with a radius of $9''$ (2.32 pixels) 
in order to exclude many nearby background stars (see Figure~\ref{f:mega} and Table~\ref{t:detect}). 
The middle panel of Figure \ref{f:lc} shows the detrended version of our extracted light-curve 
using the best-fit low-frequency model produced during the light-curve calibration.

The determination of the physical radius of the planet candidate and size of its orbit 
requires an accurate characterization of the host star, 
which we present in Section \ref{s:star}. 
In this work, we adopt the following conventions from IAU 2015 Resolution B3 
for the nominal radii for the Sun and Earth,
which we apply to convert the measured transit quantities $a/R_\star$ and $R_p / R_\star$ to 
physical and terrestrial units \citep{iau2015, iau2015b}: 
$1~\Rnom = 6.957 \times 10^8$~m and $1~\REnom = 6.3781 \times 10^6$~m, 
where this nominal terrestrial radius is Earth's ``zero tide'' equatorial value.

We modeled the light-curve with \texttt{EXOFAST} \citep{EXOFAST}.\footnote{\url{http://astroutils.astronomy.ohio-state.edu/exofast/}}\footnote{We performed preliminary modeling on a 20 hr segment of 
our detrended and phase-folded light-curve centered on the approximate 
time of transit using the web interface for \texttt{EXOFAST}, 
which simplified and sped up the fitting procedure.}
\texttt{EXOFAST} is an IDL-based transit and RV fitter for solving single-planet systems
that employs
the \citet{Mandel2002} analytic light-curve model,  
limb darkening parameters from \citet{Claret2011}, 
and accounts for the long 30 minute \textit{K2} cadence.
\texttt{EXOFAST} requires prior information on the time of transit and the period of the orbit;
the stellar temperature, metallicity, and surface gravity; 
and, without radial velocities (RVs), 
\citeauthor{EXOFAST} recommended fixing the 
orbit geometry to circular, as the light-curve 
does not provide adequate constraints on eccentricity or the argument of periastron. 

Next, we modeled the light-curve following the procedure applied 
in the Zodiacal Exoplanets In Time (ZEIT) program,
described in 
\citet{Mann2016, Mann2017, Mann2017Hyades}, 
which employs model light-curves generated with 
the BAsic Transit Model cAlculatioN code \citep[\texttt{batman};][]{batman} 
and the quadratic limb-darkening law sampling method from \citet{Kipping2013}.
We also accounted for the 30 minute cadence 
and assigned a Gaussian prior on the stellar density of $\rho_\star = 1.17 \pm 0.12~\rho_\odot$ 
derived from our estimates of the star's mass and radius.
The posterior distributions of the various model parameters
were sampled with the affine-invariant 
Markov chain Monte Carlo
(MCMC) code \texttt{emcee} \citep{emcee}.

\begin{figure*}\begin{center}
\includegraphics[width=6.5in]{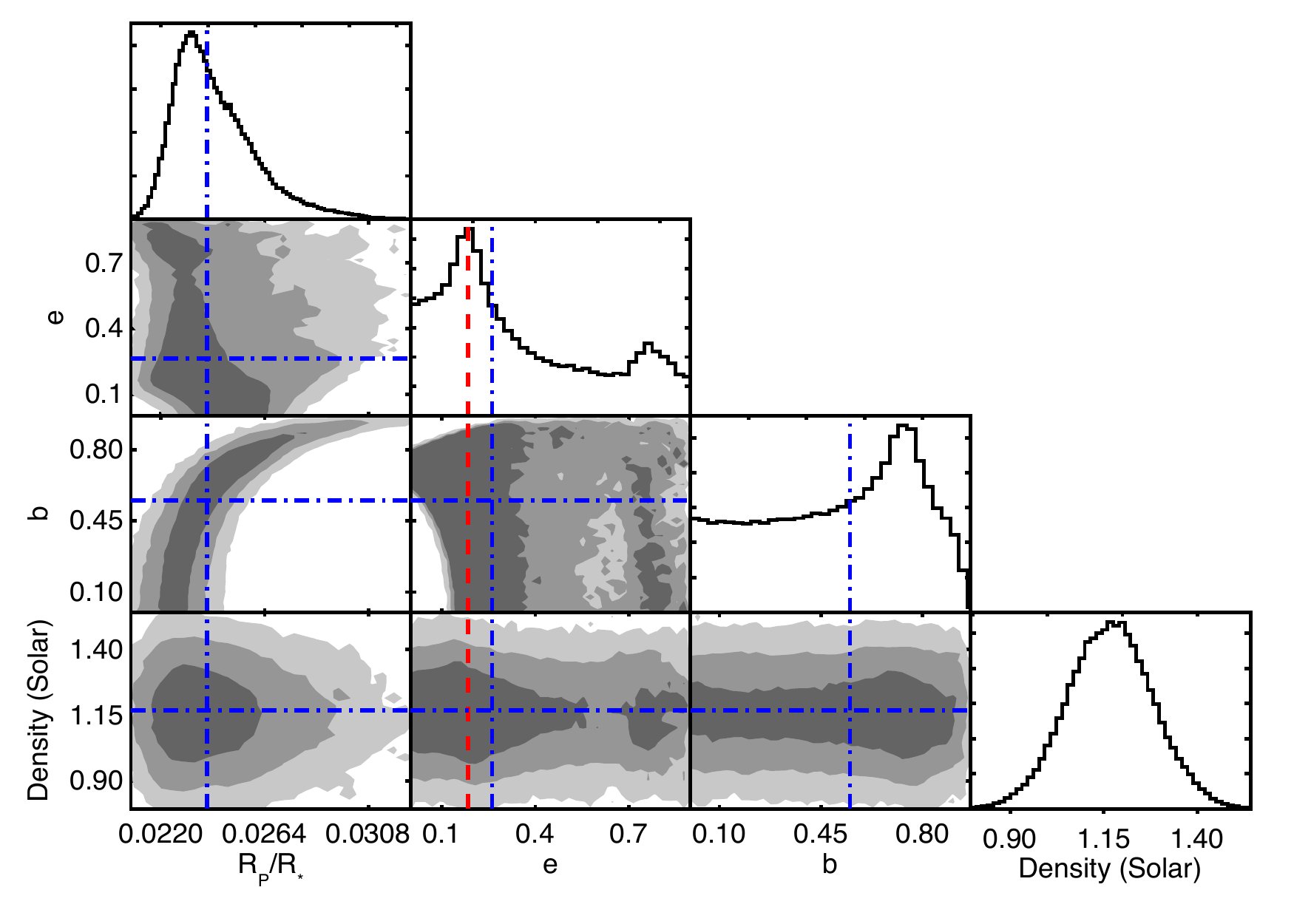}
	\caption{Results of the ZEIT MCMC transit-fitting procedure. 
	This corner plot shows posterior distributions and correlations of a subset of the
	transit-fit parameters, including 
	the ratio of planetary-to-stellar radius $R_p / R_\star$, 
	eccentricity $e$, 
	impact parameter $b$, and
	stellar density in solar units.
	The blue lines indicate the median values for each distribution; 
	the red line shows the mode for the eccentricity plot.
	The shaded regions mark the 68\%, 95\%, and 99.7\% contours of the MCMC posteriors.
	\label{f:corner}}
\end{center}\end{figure*}

In Table~\ref{t:prop}, we report the median values for each parameter 
and errors as the 84.1 and 15.9 percentile values (i.e., 1$\sigma$ for a Gaussian distribution).
Figure~\ref{f:corner} plots the posterior distributions and correlations for a subset of
transit-fit parameters resulting from our MCMC analysis.
Note that duration and inclination are not fit 
but are derived from the stellar density and impact parameter. 
The eccentricity and argument of periastron are weakly constrained, 
which is common for long-cadence data, 
especially when lacking RV data. 
Likewise, the stellar density posterior is essentially a reflection of the 
adopted prior, as it encapsulates the uncertainty in eccentricity.
Because the posteriors are not necessarily Gaussian or symmetric, 
it is possible that the median values reported here
for one set of values do not perfectly translate to that of others. 
Similarly, the plotted model is the best fit (i.e., highest likelihood), which is not necessarily the same as the median value.

The bottom panel of Figure~\ref{f:lc} shows this same light-curve 
phase-folded according to the \PerOrb~day period along with 
the model solution from 
the ZEIT procedure. 
As we will discuss later in Section \ref{s:false}, there is a star $\sim$4$''$ south of \thisstar\ and fainter by $\sim$4 mag. 
We corrected the light-curve for the dilution of the transit caused by this star by
assuming that this star contributes a flat signal with a relative flux of $\approx$1/40, 
which increases the derived radius by a few percent.
With the ZEIT procedure, we find that \thisplanet\ has a radius of $R_p = 2.48 \pm \eRplanet$~\rearth .
For comparison, \texttt{EXOFAST} returned $R_p = 2.42 \pm 0.14~\rearth$, 
which is consistent to 0.3~$\sigma$.
The \texttt{EXOFAST} uncertainty appears lower  
because we forced it to fit a circular orbit,
whereas eccentricity was allowed to float in the ZEIT procedure. \\

\textit{Note to readers of this preprint:} Our calibrated 
light-curve 
is included in the arXiv source file.

\newcommand{\bjdtdb}{\ensuremath{\rm {BJD_{TDB}}}}
\newcommand{\lsun}{\ensuremath{\,L_\Sun}}
\newcommand{\fave}{\langle F \rangle}
\newcommand{\fluxcgs}{10$^9$ erg s$^{-1}$ cm$^{-2}$}

\begin{deluxetable*}{lccc}
\tablecaption{Stellar and Planetary Properties for \thisplanet \label{t:prop}}
\tablewidth{0pt}
\tablehead{
  \colhead{Parameter} & 
  \colhead{Value}     &
  \colhead{68.3\% Confidence}     &
  \colhead{Source} \\ 
  \colhead{} & 
  \colhead{}     &
  \colhead{Interval Width}     &
  \colhead{}  
}
\startdata
\multicolumn{4}{l}{\emph{Other Designations:} \thisepic , NOMAD 0742--0804492, CWW 93,
2MASS J19162203$-$1546159} \\
\sidehead{\textit{Basic Information}}
R.A. [hh:mm:ss] & 19:16:22.04 & $\cdots$ &  Gaia DR1 \\
Decl. [dd:mm:ss]  & $-$15:46:16.37   &  $\cdots$  &  Gaia DR1 \\
Proper motion in R.A. [\ensuremath{\rm mas\,yr^{-1}}] &  $-0.5$ & 1.0  & HSOY  \\
Proper motion in decl. [\ensuremath{\rm mas\,yr^{-1}}]& $-25.0$ & 1.0 & HSOY  \\
Absolute RV [\kms] & 41.576 & $0.004 \pm 0.1$ & HARPS \\ 
$V$ magnitude & 12.71 & 0.04 & APASS\\ 
Distance to R147~[pc]& 295 & 5 & C13\\
Visual extinction ($A_V$) for R147 [mag] & 0.25   & 0.05 & C13\\ 
Age of R147 [Gyr] & 3 &  0.25 & C13 \\
\sidehead{\textit{Stellar Properties}}
$M_\star$~[$M_\odot$] & \StarMass & \eStarMass     & Phot+Spec+Iso \\
$R_\star$~[$R_\odot$] & \StarRadius & \eStarRadius & Phot+Spec+Iso \\
$\log g_\star$~[cgs] & \StarLogg & \eStarLogg & Phot+Spec+Iso \\
$T_{\rm eff}$, adopted [K]      & \StarTeff & \eStarTeff & Phot+Spec+Iso\\
Spectroscopic metallicity      & $+0.14$ & 0.04 & SME \\
R147 metallicity & $+0.10$ & 0.02 & SME \\
$v\sin{i}$ [\kms]               &  \SpecVsini     & \eSpecVsini        & SME\\
Mt. Wilson $S_{\rm HK}$         & 0.208     & 0.005      & Section~\ref{s:hk}  \\
Mt. Wilson $\log{R'_{\rm HK}}$  & $-4.80$   & 0.03       & Section~\ref{s:hk}  \\
\sidehead{\textit{Planet Properties}}
Orbital period, $P$~[days]      & \PerOrb   & \ePerOrb   & Transit \\
Radius ratio, $R_P/R_\star$     & \RpRs     & \eRpRs     & Transit \\
Scaled semimajor axis, $a/R_\star$ & \ScaledSemi & \eScaledSemi & Transit \\
Transit impact parameter, $b$   & \Impact   & \eImpact   & Transit \\
Orbital inclination, $i$~[deg]  & \Incl     & \eIncl     & Derived \\
Transit Duration, $t$~[hr] & \Tdur  & \eTdur     & Derived \\
Time of Transit $T_{0}$~[BJD$-$2,400,000]   & \TransitTime & \eTransitTime & Transit\\ 
Planet radius $R_P$~[\rearth]   & \Rplanet  & \eRplanet  & Converted \\
\enddata
\tablecomments{Coordinates are from \textit{Gaia} DR1 \citep{GaiaDR1a}; 
proper motions are from HSOY \citep{HSOY}; 
the RV is the weighted mean for the six HARPS RVs and 
the uncertainties represent the precision and accuracy, respectively, 
where the accuracy is an approximation of the uncertainty in the IAU absolute velocity scale (Table~\ref{t:rvs});
the $V$ magnitude is from APASS \citep{APASS9}; 
the distance, age, and extinction are from \citet{Curtis2013}; 
the cluster metallicity was derived from SME analysis \citep{sme} of seven solar analog members of R147 \citep{Curtis2016PhD}; 
the metallicity and projected rotational velocity were derived from SME analysis of the MIKE spectrum;
the adopted stellar mass, radius, temperature, and surface gravity were derived by analyzing 
the available spectroscopic and photometric data together with isochrone models (see Section~\ref{s:star});
and the transit parameters are the median values and the 68\% interval 
from the posterior distributions resulting from our MCMC analysis, 
except for the transit duration and inclination, which are derived from 
from the stellar density and impact parameter.
The planetary radius, measured relative to the stellar radius, is converted to terrestrial units using 
values for the Earth and Sun radius from IAU 2015 Resolution B3.
Chromospheric activity indices were measured from 
Hectochelle spectra following principles described in \citet{wright2004}.}
\end{deluxetable*}

\section{Properties of the host star} \label{s:star}

\citet{Curtis2013} demonstrated that \thisstar\ is a member of Ruprecht 147, 
and therefore it should share the properties common to the cluster, 
including 
a spectroscopic metallicity of [Fe/H] = +0.10~dex 
\citep{Curtis2016PhD},
an an age of 3 Gyr,
a distance of 295~pc based on the 
distance modulus of $m - M = 7.35$,
and an interstellar extinction of $A_V = 0.25$ mag, 
derived from fitting Dartmouth 
isochrone models \citep{dartmouth} 
to the optical and NIR color--magnitude diagrams (CMDs).
We estimate the mass and radius of this star with a combination 
of spectroscopic and photometric data 
and then argue that it is likely single (i.e., not a stellar binary). 

\subsection{Spectroscopy}

On 2016 July 15, we used the MIKE spectrograph \citep{MIKE} 
on the 6.5~m Magellan Clay Telescope 
at Las Campanas Observatory in Chile to acquire a spectrum 
of \thisstar\ with the 0\farcs70 slit, 
corresponding to a spectral resolution of $R=42,000$; 
the per-pixel signal-to-noise ratio is 
$S/N=130$ and 208 at the peaks of the \mgb\ and 5940--6100~\AA\ orders, respectively. 
We also observed six other solar analogs in R147 at $R = 42,000$ and 20 solar analogs in the field at 
$R = 55,000$
(including 18~Sco and the Sun as seen from the reflection off of the dwarf planet Ceres, 
which we observed with both resolution settings).
We reduced these spectra with the Carnegie Python pipeline 
(``CarPy''),\footnote{\url{http://code.obs.carnegiescience.edu/mike}}
which performs the standard calibrations 
(i.e., overscan, bias, flat-field, sky-background, and 
scattered-light corrections, 
and mapping in wavelength using thorium--argon lamp spectra). 

We analyzed these spectra with version 423 of Spectroscopy Made Easy \citep[SME;][]{sme} 
following the \citet{valenti2005} procedure. 
Adopting stellar properties for the field stars from \citet{Brewer2016},
that sample spans 
\teff\ = 5579--5960~K, 
\logg\ = 4.10--4.50~dex, and
[Fe/H] = $-0.09$ to $+0.14$~dex.
We find median offsets and standard deviations between the \citet{Brewer2016} properties and our values of 
\teff\ =~$-11$,~27~K, 
\logg\ =~$-0.04$, 0.035~dex, and 
[Fe/H]~=~$-0.016$, 0.02~dex. 
These numbers illustrate our ability to reproduce the \citet{Brewer2016} results with 
different data and a different SME procedure 
\citep[we do not employ the expanded spectral range and line list of][]{Brewer2015,Brewer2016}, 
and they are all within the SME statistical uncertainties quoted by \citet{valenti2005} of 
44~K, 
0.06~dex, and 
0.03~dex, respectively.

Regarding the sample of seven solar analogs in R147, 
after applying the offsets, 
we find [Fe/H]$ = +0.10 \pm 0.04$~dex, 
\added{where the uncertainty is the standard deviation
of the sample; 
the standard deviation of the mean is $\pm$0.02~dex and is reported in Table~\ref{t:prop}.} 
While the R147 dispersion is higher than that measured in the field star sample 
relative to the \citet{Brewer2016} metallicities, 
this is probably due to the typically lower $S/N$s and spectral resolutions 
of the R147 spectra (the stars are much fainter) compared to the field stars taken from \citet{Brewer2016}, 
and not intrinsic to the sample.

For a separate project, Iv\'{a}n Ram\'{i}rez measured stellar properties for five of these solar analogs with 
the same or similar MIKE spectra (since his work, we have collected higher-quality data for particular stars for our analysis described here). 
Following \citet{Ramirez2013},
he employed a differential analysis with respect to the Sun by enforcing the 
excitation/ionization balance of iron lines
using the MOOG spectral synthesis code.\footnote{\url{http://www.as.utexas.edu/∼chris/moog.html}}
He also fit the telluric-free regions of the wings of H$\alpha$ using the \citet{Barklem2002} grid.
For the same project, Luca Casagrande measured IRFM temperatures for these stars following \citet{Casagrande2010}.
For these five stars, we find a median offset and standard deviation for our SME values minus theirs of
$-26 \pm 29$~K for the Fe method, $-4 \pm 22$~K for H$\alpha$, and $-31 \pm 73$~K for IRFM 
(I. Ram\'{i}rez \& L. Casagrande 2013, private communication).
These differences are all within the uncertainties quoted and cross-validate our adopted temperature scale. 

Based on our results for the field star sample, 
the R147 members, 
and the SME statistical uncertainties quoted by \citet{valenti2005},
we adopt the following spectroscopic parameter precisions: 
50~K for \teff, 
0.06 for \logg , and 
0.04~dex for [Fe/H].
Our error analysis assumes that our uncertainties are limited by 
the data quality and our analysis technique, 
and not systematics inherent in the models. 
As our sample is comprised of stars quite similar to the Sun, 
the issues that tend to plague analyses of non-solar-type stars are assumed to be largely mitigated. 
The procedure accurately reproduces the Sun's properties by design, 
as the line data were tuned to the solar spectrum; therefore, 
we assume that it can safely be applied to solar twins with 
the same degree of accuracy, 
and we adopt our precision estimates as our total parameter uncertainties.

For \thisstar, we found an 
effective temperature of \teff\ =  5697~K,
surface gravity of \logg\ = 4.453~dex,
iron abundance of [Fe/H] = +0.141~dex,
and rotational broadening of \vsini~=~1.95~\kms\  
when we adopted the macroturbulence relation from \citet{valenti2005} (i.e., $v_{\rm mac} = 3.87$~\kms ).
Adopting our preferred parameters for the Dartmouth isochrone model to describe the R147 cluster 
(age of 3 Gyr and [Fe/H] = +0.1~dex) 
and querying the model at the spectroscopic temperature yields 
an isochrone-constrained surface gravity of $\logg = 4.483$~dex, 
which we adopt for $\logg$.
We refit the spectrum with metallicity fixed to the cluster value and \logg\ fixed to this isochrone value, 
which returned $\teff = 5672$~K and $\vsini = 1.3$~\kms ,
which is only 25~K cooler than the unconstrained fit.

\subsection{Stellar mass and radius}
\begin{deluxetable}{lcccc}
\tabletypesize{\scriptsize}
\tablecaption{Photometry for \thisstar \label{t:phot}}
\tablewidth{0pt}
\tablehead{
\colhead{Instrument} & \colhead{Band} & 
\colhead{mag} & \colhead{error} & \colhead{$A / A_V$} \\
}
\startdata
\textit{Gaia} & $G$   & 12.46 & $\cdots$  & 0.861 \\
APASS         & $B$   & 13.50 & 0.03 & 1.297 \\
APASS         & $V$   & 12.71 & 0.04 & 1.006 \\
CFHT/MegaCam  & $g'$  & 13.02 & 0.02 & 1.167 \\
APASS         & $g$   & 13.07 & 0.01 & 1.206 \\
CFHT/MegaCam  & $r'$  & 12.46 & 0.02 & 0.860 \\
APASS         & $r$   & 12.47 & 0.07 & 0.871 \\
CFHT/MegaCam  & $i'$  & 12.27 & 0.02 & 0.656 \\
APASS         & $i$   & 12.26 & 0.04 & 0.683 \\
2MASS         & $J$   & 11.29 & 0.02 & 0.291 \\
2MASS         & $H$   & 11.00 & 0.03 & 0.184 \\
2MASS         & $K_S$ & 10.86 & 0.02 & 0.115 \\
UKIRT/WFCAM   & $J$   & 11.30 & 0.02 & 0.283 \\
UKIRT/WFCAM   & $K$   & 10.92 & 0.02 & 0.114 \\
\textit{WISE} & $W1$  & 10.75 & 0.02 & 0.071 \\
\textit{WISE} & $W2$  & 10.84 & 0.02 & 0.055 \\ 
\enddata
\tablecomments{(1) Name of instrument or survey. 
(2) Photometric band/filter employed. 
(3,4) Magnitude and uncertainty for that observation,
where pipelines/surveys quoted errors below 0.01~mag, we set the value to 0.02~mag for analysis.
(5) Interstellar reddening coefficients computed by the Padova/PARSEC isochrone group \citep{parsec} 
for a G2V star using the \citet{Cardelli1989} extinction law and following a procedure similar to that
described by \citet{Girardi2008}.} 
\end{deluxetable}

We estimated the mass and radius of \thisstar\ by combining our spectroscopic results 
with the optical and NIR photometry provided in Table \ref{t:phot}. 
We assembled photometry from 
\textit{Gaia} \citep{GaiaDR1a, GaiaDR1b},
the AAVSO Photometric All-Sky Survey \citep[APASS;][]{APASS9}
the CFHT's MegaCam \citep{MegaCam} presented by \citet{Curtis2013},
the Two Micron All-Sky Survey \citep[2MASS;][]{2MASS},
the United Kingdom Infra-Red Telescope's (UKIRT)
Wide Field Infrared Camera \citep[WFCAM;][]{WFCAM} that was acquired by 
coauthor A.L.~Kraus in 2011 and accessed from the WFCAM Science Archive,\footnote{\url{wsa.roe.ac.uk}}
and 
NASA's \textit{Wide-field Infrared Survey Explorer} \citep[\textit{WISE};][]{WISE}.

First, we used the PARAM 1.3 input form---the ``web interface for the Bayesian estimation of stellar parameters'' 
described by \citet{param}---to estimate the mass and radius of the 
host star.\footnote{\url{http://stev.oapd.inaf.it/cgi-bin/param\_1.3}}  
This service uses the PARSEC stellar evolution tracks \citep[version 1.1;][]{parsec}. 
The procedure requires as input the effective temperature, metallicity, 
parallax, and $V$ magnitude.
We adopted the \citet{Curtis2013} distance modulus and visual extinction to estimate the 
dereddened magnitude ($V_0 = V -0.25 = 12.458$) 
and parallax of $\pi = 3.39$~\mas\ 
(calculated from 295~pc).\footnote{The 
cluster-averaged parallax from the \textit{Tycho--Gaia} Astrometric Solution 
\citep[TGAS;][]{TGAS} from \textit{Gaia} DR1 \citep{GaiaDR1a, GaiaDR1b}
is consistent with this at 3.348~\mas, translating to 299~pc, 
based on 33 RV and AO single members \citep{Curtis2016PhD}.}
For parameter uncertainties, we adopted 50~K and 0.05~dex for \teff\ and [Fe/H], 
and 0.05~mag for $V_0$ and 0.15~mas for parallax based on the uncertainty in 
$A_V$ and $m - M$.
PARAM 1.3 returned 
age $t_\star = 1.7 \pm 1.6$~Gyr, 
mass $M_\star = 1.009 \pm 0.027$~\msun,
$\logg_\star = 4.474 \pm 0.029$~dex~(cgs), and
radius $R_\star = 0.934 \pm 0.029$~\rsun .

Next, we estimated the mass and radius using the 
Python \texttt{isochrones} package \citep{Morton2015}.\footnote{\url{https://github.com/timothydmorton/isochrones}}
We adopted the spectroscopic \teff\ and  \logg\ values,
the cluster metallicity and parallax,
and the de-reddened broadband photometry from Table \ref{t:phot}, 
and ran the fit assuming the photometry was derived from a blended and physically associated binary. 
Only 56\% of nearby field stars are single \citep{Raghavan2010}, 
so it is important to consider at least binarity when characterizing this system 
(\citeauthor{Raghavan2010} also found that 11\% of nearby stars are in $3+$ multiples).
We used grid models from the Dartmouth Stellar Evolution Database \citep{dartmouth} 
and sampled the posteriors using MultiNest \citep{Multinest1, Multinest2, Multinest3}
implemented in Python with the \texttt{PyMultiNest} package \citep{PyMultiNest}.
Expressing uncertainties as the 68.3\% (1$\sigma$) confidence intervals
of the posterior distributions,
we found
$M_1 = 1.013 \pm 0.016$~\msun ,
$R_1 = 0.944 \pm 0.021$~\rsun ,  and
$M_2 = 0.238 \pm 0.104$~\msun .

If the host is indeed single, then we can expect the parallax-constrained photometric analysis
to return a small secondary mass with a value at approximately the 
threshold where its contributed flux is on par with the photometric errors (i.e., consistent with no secondary).
Based on this low secondary-mass estimate, 
there is no evidence from the photometry for a secondary companion: 
the difference in magnitude between the resulting primary and secondary stars
is $\Delta V = 7.29$ and $\Delta K = 4.18$, 
which is too large of a contrast to detect from these data
(i.e., the difference between the primary and the combined magnitude of both stars is 
0.001 mag in $V$ and 0.023 mag in $K$, the latter of which is on par with 
the measurement errors). 
We reran the fit with \texttt{isochrones} assuming a single star, 
which returned 
$M_\star= 1.004 \pm 0.017$~\msun ,
$R_\star = 0.938 \pm 0.022$~\rsun ,
$d = 302 \pm 8$~pc,
$A_V = 0.29 \pm 0.05$, and
$t = 2.5 \pm 1$~Gyr.
The age, distance, and reddening values are consistent with 
the CMD isochrone fitting results from \citet{Curtis2013}; 
the mass and radius is consistent with the PARSEC/PARAM result quoted above.

To further test possible systematics in the isochrone fitting methods and models,
we derived stellar properties using the 
\texttt{isoclassify} code \citep{isoclassify},\footnote{\url{https://github.com/danxhuber/isoclassify}} 
conditioning spectroscopic \teff, \logg, \feh, parallax, and 2MASS $JHK$ photometry on a grid of interpolated MIST isochrones \citep{MIST}. 
This returned 
$M_\star = 1.014 +0.021 -0.022$~\msun, 
$R_\star = 0.960 +0.027 -0.024$~\rsun, 
$d = 309 +9 -9$~pc, 
$A_V = 0.09 +0.27 -0.24$~mag, and 
$t = 2.3 +1.6 -1.3$~Gyr, 
in excellent agreement with the values derived from other isochrone models and methods.

Again, systematic uncertainties in the models are likely negligible due to the 
Sun-like nature of the host star 
\citep[whereas, for example, K-dwarf models are known to diverge
between PARSEC and Dartmouth;][]{Huber2016, Curtis2013}.
The dispersion in masses and radii derived from the three isochrone models 
are well within the uncertainties returned by each method, 
so we adopt the maximum uncertainties from the various experiments 
as our final measurement uncertainties and we take the mean as our final values: 
$M_\star = 1.009 \pm 0.027$~\msun\ and $R_\star = 0.945 \pm 0.027$~\rsun.

According to the MIST model, a 3 Gyr star with 
mass $M_\star = 1.009$~\msun\ and [Fe/H] = +0.1~dex has 
\teff\ = 5695~K. 
This value is only 2~K cooler than our SME result, 
and so we adopt this value as the effective temperature of this star.

\subsection{\thisstar\ Is Likely Single}\label{s:sin}

\added{It is important to test \thisstar\ for stellar multiplicity. 
We need to know if it is a binary or higher-order multiple so 
we can confidently assume which star hosts the transiting object and
how much the light from the companion(s) 
has diluted the observed transits.
We assembled a variety of observational evidence, 
outlined below,
that collectively indicates that \thisstar\ is likely single. 
The various constraints derived from these data 
are summarized in Figure~\ref{f:binary}, 
which shows the parameter space for a range of binary scenarios with 
secondaries described by $K$-band contrast (left axis) and 
isochrone-estimated stellar mass (right axis) as a function 
of projected separation in angular units (bottom axis; out to 1000~mas) and 
physical units (top axis; out to 300~AU).}\\

\noindent \textbf{Photometry:} 
Reiterating our result from the previous subsection, 
modeling the broadband photometry with the \texttt{isochrones} package 
suggests that \thisstar\ does not have a companion 
with a mass $M_2 > 0.34$~\msun. 
\added{Such a secondary 
would be at least 
$\sim$321 times fainter than the primary in $V$;
correcting for transit dilution would only increase the transit depth by 0.3\% and the planet radius by 0.15\%
Basically, the effect of any
binary companion allowed by the photometric 
modeling is negligible.
This constraint is illustrated in Figure~\ref{f:binary} 
by the light blue shaded region at the top.} \\

\noindent \textbf{Adaptive optics imaging and coronagraphy:} 
We acquired natural guide star AO imaging in $K'$ 
($\lambda = 2.124~\mu$m)
with NIRC2 on the Keck II telescope. 
We also used the ``corona600'' occulting spot, 
which has a diameter of 600~mas 
and an approximate transmission of 0.22\% in $K'$.
The observations were acquired, reduced, and analyzed following \citet{Kraus2016}. 
Table~\ref{t:limits} lists the $K'$ detection limits 
as a function of angular separation from \thisstar\ 
ranging from 150 to 2000~mas. 

\begin{figure}\begin{center}
\includegraphics[width=4.5in, trim=3.5cm 0 0 1.2cm]{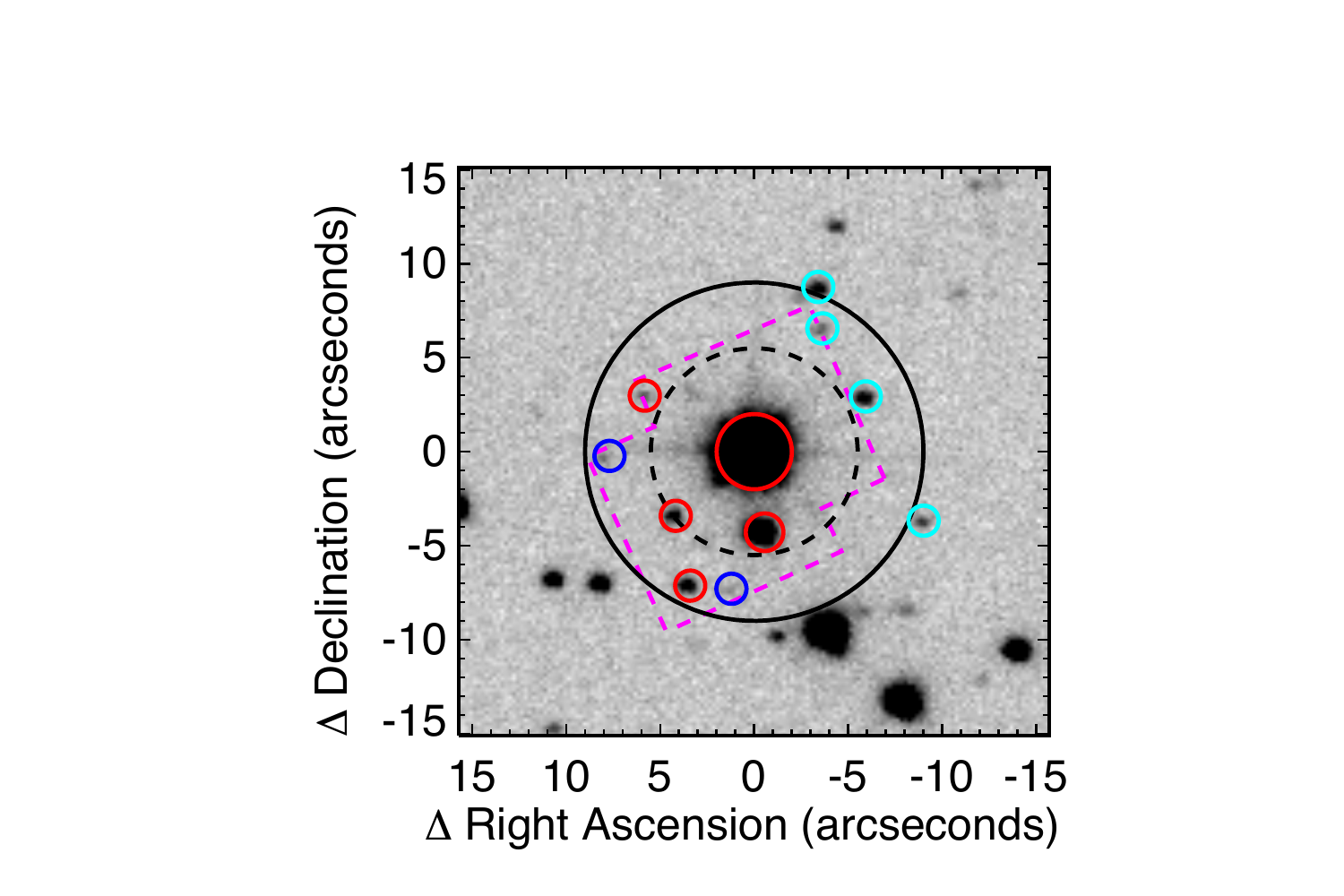}
	\caption{Image of \thisstar\ and neighboring stars from 
	UKIRT/WFCAM, taken in 2011.
	The solid black circle shows the $9''$ radius aperture used to extract the 
	light-curve. The dashed black circle has a radius of $5.5''$ and is the smallest 
	aperture we tested; the transits are still visible, which means 
	that the object is either transiting \thisstar\ or the fainter star 4$''$ southward.
	The dashed magenta line traces out the Keck II NIRC2 footprint:  
	six stars were detected, four of which cross-matched with the UKIRT catalog (red) 
	and two of which were apparently too faint, though they show some signal in the image (blue). 
	Four other stars are detected in the WFCAM image within 10$''$ but were missed 
	by NIRC2 due to the size, placement, and orientation of the field (cyan).
	Properties of these 10 neighboring stars are listed in Table~\ref{t:detect}.
	\label{f:mega}}
\end{center}\end{figure}

\begin{figure}\begin{center}
\includegraphics[width=3.75in,trim=2cm 0 0 0]{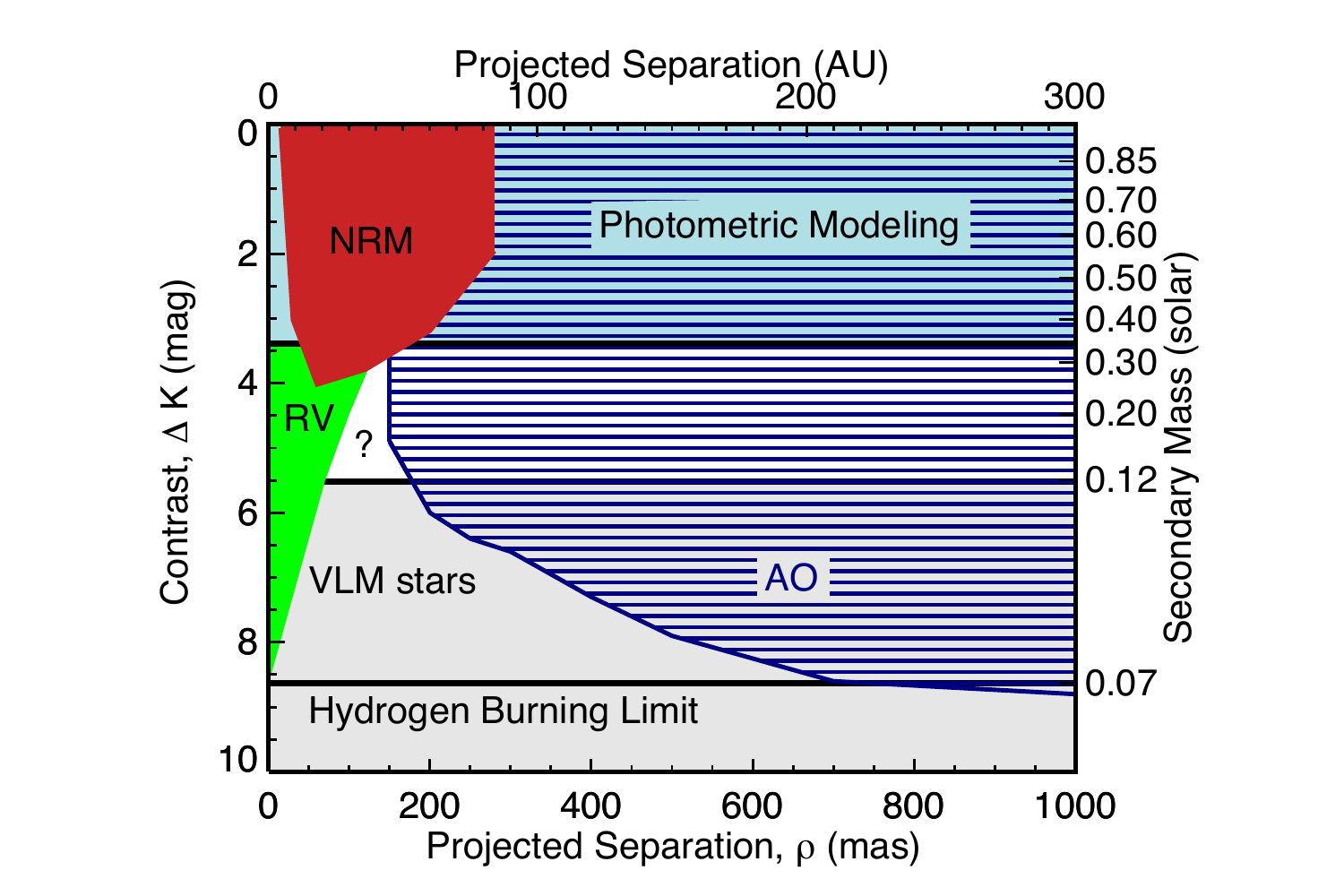}
	\caption{Constraints on binary companionship for 
	hypothetical secondaries with 
	$K'$-band contrast (left) or stellar mass (\msun; right) 
	as a function of 
	projected separation in angular units (mas; bottom) or 
	physical units (AU; top).
	At separations of $\rho > 200$~mas, the NIRC2 AO imaging and coronagraphy 
	(dark blue lined region)
	probe deeper than the very low mass stars and reaches down to the
	hydrogen-burning limit at $\sim$700~mas (gray shaded region), 
	which is useful for searching for background blends; 
	the NIRC2 non-redundant masking data reach closer to the primary star, 
	but not quite as deep (red shaded region). 
	Modeling the broadband photometry with \texttt{isochrones} rules out 
	secondaries of any separation with masses greater than $M_2 \ga 0.34$~\msun\ 
	(light blue shaded region). 
	Combining these various constraints leaves a small region of
	parameter space under 45~AU (projected) for systems with $M_2 \lesssim 0.34$~\msun.
	The precise HARPS RVs can rule out much of this remaining parameter space for edge-on orbits 
	(green shaded region); 
	accounting for possible inclination of the 
	binary orbital plane relative to the 
	primary--planet orbit will 
	restrict this to smaller separations. 
	\label{f:binary}}
\end{center}\end{figure}

Table~\ref{t:detect} lists six stars within $8''$ that were detected, 
including coordinates; 
angular separation, position angle, and $K'$ contrast relative to \thisstar;
and photometry from 
\textit{Gaia}, 
CFHT/MegaCam, 
and UKIRT/WFCAM.
This table also lists four stars within $10''$ detected in the UKIRT imaging 
that were missed by NIRC2. 
Figure~\ref{f:mega} shows a $30''$-square $K$-band 
image from UKIRT/WFCAM centered on the host star 
and highlights the noncoronagraphic imaging footprint 
(magenta dashed line); 
note that we had to offset the pointing after the first image in
order to get the bright neighboring star onto the detector, 
which is why there is effectively a double footprint.
For reference, two circles with radii of 5\farcs5 and $9''$ are also 
overlaid to show the approximate extraction apertures used 
to produce light-curves from the \textit{K2} data.
The AO imaging and coronagraphy yielded six detections, 
four of which were matched in the UKIRT imaging (red circles), 
and two of which were apparently fainter than the UKIRT source catalog limit (blue circles), 
but nevertheless show up in the image.
Due to the placement, size, and orientation of the NIRC2 footprint, 
four stars within $10''$ of the host were missed 
but show up in WFCAM (cyan circles).

We calculated proper motions for the eight stars that matched in both \textit{Gaia} and either or both 
MegaCam and WFCAM and found that none but the final entry appear comoving with R147. 
We also inspected optical and NIR CMDs with the cluster Dartmouth model overlaid 
and noted that stars 1, 3, 8, and 9 are inconsistent with membership, 
whereas 6, 7, and 10 appear near but beyond the base of the Dartmouth isochrone.
As 6 and 7 appear to be ruled out by their discrepant proper motions, 
this leaves 10 as the sole candidate member in this list.
Although too faint for \textit{Gaia}, 
it is conceivable that we could measure its proper motion with 
additional NIRC2 images in the future: 
the uncertainty on $\rho$ is under 5~mas, whereas R147 moves at $-28$~\mas\ in declination, 
so two observations spaced approximately by one year should clearly reveal any comoving stars while
canceling out the parallax effect.

Only two stars are detected within 5\farcs5, 
which is the radius of the smallest circular moving aperture that we used to 
extract light-curves. 
One star is near the edge of this radius and is nearly 480 times fainter 
than \thisstar . 
The other, at 4\farcs2 southward, is 40 times fainter,
and we consider it our primary false-positive source. 

These constraints are illustrated in Figure~\ref{f:binary} by the 
dark blue shading, which covers the majority of the upper right region.
Masses/contrasts below the hydrogen-burning limit at $\sim$0.07~\msun\ are 
shaded gray and found below the black horizontal line toward the bottom of the figure, which the AO limit reaches at $\sim$700~mas---this depth 
is not only important for searching for stellar binaries, 
but also for identifying faint, unassociated stars in the background.
The lowest mass star represented in the Dartmouth isochrone model 
is $M \approx 0.12$~\msun: we also shade this region gray and label it ``VLM'' for 
``very low mass star''
to distinguish it from the region below the substellar boundary 
while highlighting that this represents a small region of the secondary mass parameter space
compared to the top-half of the figure.
\\

\noindent \textbf{Keck/NIRC2 aperture-masking interferometry:} 
We also acquired nonredundant aperture-masking interferometry data
for \thisstar\ on 2017 June 22 in natural guide star mode, 
along with EPIC~219511354 for calibration.
For the target and reference star, 
we obtained four (three) interferograms for a total of 80 (60)~s on \thisepic\ (EPIC~219511354),
which we analyzed following
\citet{Kraus2008AO, Kraus2011, Kraus2016}. 
We report no detections within the limits quoted in 
Table~\ref{t:apmask}.
These constraints are illustrated in Figure~\ref{f:binary} by the red shaded region, 
which is drawn according to the midpoints of the angular separation ranges 
listed in Table~\ref{t:apmask}.\\

\begin{deluxetable*}{cccccccccccccc}
\tabletypesize{\footnotesize}
\tablewidth{0pt}
\tablecaption{Keck/NIRC2 Imaging Detection Limits \label{t:limits}}
\tablehead{
\colhead{MJD} & \colhead{Filter +}    & \colhead{Number of} & \colhead{Total}         & 
   \multicolumn{10}{c}{Contrast Limit ($\Delta K'$ in mag) at Projected Separation ($\rho$ in mas) } \\
\colhead{}    & \colhead{Coronagraph} & \colhead{Frames}    & \colhead{Exposure (s)} & 
\colhead{150} & \colhead{200} & \colhead{250} & \colhead{300} & 
\colhead{400} & \colhead{500} & \colhead{700} & \colhead{1000} & \colhead{1500} & \colhead{2000} 
}
\startdata
57933.42 &   $K'$     &  6 &  120.00 &  4.9 &  6.0 &  6.4 &  6.6 &  7.3 &  7.9 &  8.6 &  8.8 &  8.9 &  8.9 \\ 
57933.43 &   $K'$+C06 &  4 &   80.00 &  $\cdots$  &  $\cdots$  &  $\cdots$  & $\cdots$ &  7.2 &  7.2 &  7.9 &  9.3 &  9.7 &  9.8 \\ 
\enddata
\tablecomments{The second entry is for the coronagraphic imaging observations, which obstructs the inner 3~mas radius.}
\end{deluxetable*}

\begin{deluxetable*}{rccccccccccc}
\tabletypesize{\footnotesize}
\tablewidth{0pt}
\tablecaption{Keck/NIRC2\tablenotemark{a} and UKIRT/WFCAM Detected Neighbors \label{t:detect}}
\tablehead{
\colhead{\#} & \colhead{R.A.} & \colhead{Decl.} & \colhead{$\rho$} & \colhead{PA} & \colhead{$\Delta K'$} & 
\colhead{$G$} & \colhead{$g'$} & \colhead{$r'$} & \colhead{$i'$} & \colhead{$J$} & \colhead{$K$} \\ 
\colhead{} &\colhead{J2000} & \colhead{J2000} & \colhead{(mas)} & \colhead{(deg)} & \colhead{(mag)} & 
\colhead{(mag)} & \colhead{(mag)} & \colhead{(mag)} & \colhead{(mag)} & \colhead{(mag)} & \colhead{(mag)} 
}
\startdata
1 & 19:16:22.005 & $-$15:46:20.58 & 4179.9 $\pm$ 1.7 &  186.488 $\pm$ 0.023 &  4.032 $\pm$0.003 & 16.52 & 17.12 & 16.46 & 16.24 & 15.18 & 14.76 \\ 
2 & 19:16:22.319 & $-$15:46:19.68 & 5182.3 $\pm$ 2.0 &  129.033 $\pm$ 0.021 &  6.708 $\pm$0.017 & 18.84 & $\cdots$ & $\cdots$ & $\cdots$ & 17.66 & 17.25 \\ 
3 & 19:16:22.424 & $-$15:46:13.21 & 6429.6 $\pm$ 2.4 &   60.404 $\pm$ 0.020 &  7.243 $\pm$0.117 & $\cdots$ & $\cdots$ & $\cdots$ & $\cdots$ & 18.92 & 18.27 \\ 
4 & 19:16:22.118 & $-$15:46:23.57 & 7388.2 $\pm$ 3.8 &  170.104 $\pm$ 0.029 &  8.216 $\pm$0.064 & $\cdots$ & $\cdots$ & $\cdots$ & $\cdots$ & $\cdots$ & $\cdots$ \\  
5 & 19:16:22.551 & $-$15:46:16.51 & 7693.7 $\pm$ 4.4 &   91.145 $\pm$ 0.032 &  8.521 $\pm$0.076 & $\cdots$ & $\cdots$ & $\cdots$ & $\cdots$ & $\cdots$ & $\cdots$ \\ 
6 & 19:16:22.269 & $-$15:46:23.41 & 7739.2 $\pm$ 2.3 &  154.535 $\pm$ 0.015 &  6.677 $\pm$0.015 & 20.06 & 22.86 & 20.81 & 20.11 & 17.99 & 17.18 \\ \hline
7 & 19:16:21.649 & $-$15:46:13.25 & 6585.7 & 297.421 & $\cdots$ & $\cdots$      & 23.48 & 21.73 & 20.71 & 18.01 & 17.21 \\
8 & 19:16:21.807 & $-$15:46:09.67 & 7466.9 & 332.356 & $\cdots$ & $\cdots$      & 21.05 & 20.64 & 20.13 & 18.75 & 18.46 \\ 
9 & 19:16:21.821 & $-$15:46:07.48 & 9386.6 & 339.730 & $\cdots$ & 18.896 & 19.07 & 18.62 & 18.32 & 17.54 & 17.06 \\ 
10 & 19:16:21.439 & $-$15:46:20.08 & 9760.9 & 247.148 & $\cdots$ & $\cdots$      & 24.31 & 23.47 & 21.71 & 18.94 & 18.17 \\
\enddata
\tablecomments{The third object was only detected in the coronagraphic observation because it fell 
on the edge of the NIRC2 imaging footprint; see Figure~\ref{f:mega}.
The objects in the lower section were detected with UKIRT but missed by NIRC2 due to the placement, 
size, and orientation of the NIRC2 field. Star 10 is the only neighbor that appears co-moving with R147 (and therefore the planet host; stars 4 and 5 were 
only detected in NIRC2 and so lack a second astrometric epoch needed to calculate proper motions), 
with a CFHT$-$UKIRT proper motion of $(\mu_\alpha \cos \delta , \mu_\delta) = (3, -31)$~\mas , 
although the baseline is relatively short at $\sim$3 years and 
we have not quantified the accuracy or precision with tests of anything near that faint.
}
\tablenotetext{a}{The relative astrometry for the NIRC2 observations was computed with the 
plate scale and rotation adopted from \citet{Yelda2010}.}
\end{deluxetable*}

\begin{deluxetable*}{lccccccc}
\tabletypesize{\footnotesize}
\tablewidth{0pt}
\tablecaption{Keck/NIRC2 aperture-masking interferometry detection limits \label{t:apmask}}
\tablehead{
\colhead{Confidence} & \colhead{MJD} & \multicolumn{6}{c}{Contrast Limit ($\Delta K'$ in mag) at Projected Separation ($\rho$ in mas) } \\
\colhead{Interval} & \colhead{} & \colhead{10-20} & \colhead{20-40} & \colhead{40-80} & \colhead{80-160} & 
\colhead{160-240} & \colhead{240-320}
}
\startdata              
99.9\%      & 57933.4 &  0.06  &   3.02 &   4.02  &    3.79  &  3.19      &   1.96 \\ 
99\% only   & 57933.4 &  0.26  &   3.24 &   4.20  &    3.97  &  3.42      &   2.2  \\
\enddata
\end{deluxetable*}

\noindent \textbf{Spectroscopy:} We observed \thisstar\ on 2017 June 2 (near quadrature, according to the transit ephemeris) 
with the High Resolution Echelle Spectrometer 
\citep[HIRES;][]{HIRES}
on the 10 m telescope at Keck Observatory. 
No secondary spectral lines were found down to 1\% of the brightness of the primary ($\sim$0.49~\msun; already ruled out by photometric modeling), 
excluding the range of under $\pm$10~\kms\ separation from the primary \citep{Kolbl2015}.\\

\noindent \textbf{RV variability:} 
We collected RVs every few years beginning in 2007, 
which show no trend due to a stellar companion 
over the baseline of nearly ten years.  
These include observations with the Lick/Hamilton and MMT/Hectochelle spectrographs 
presented in \citet{Curtis2013}, 
the HIRES spectrum mentioned above \citep{chubak2012},
and the Magellan/MIKE spectra discussed earlier.\footnote{Barycentric velocities were  
calculated with the IDL code \texttt{BARYCORR} \citep{barycenter}; 
see also \url{http://astroutils.astronomy.ohio-state.edu/exofast/barycorr.html}} 

Separately, a team led by PI Minniti targeted R147 with the 
High Accuracy Radial velocity Planet Searcher \citep[HARPS;][]{HARPS}
in 2013-2014 to look for exoplanets in R147 with masses 
greater than or approximately equal to Neptune in relatively short-period orbits 
and acquired six RV epochs with individual precisions of $\approx$10~\mps.\footnote{ESO program 
091.C-0471(A) and 095.C-0947(A), ``Hunting Neptune mass planets in the nearby old, metal rich open cluster: Ruprecht 147.''}
Data were reduced and RVs extracted with the HARPS Data Reduction Software. 
We downloaded the reduced data, including the pipeline RVs and uncertainties, 
from the ESO archive.\footnote{Values taken from the ``*ccf\_G2\_A.fits'' files.}\footnote{\url{http://archive.eso.org/wdb/wdb/adp/phase3_spectral/query}}

We recalculated the RVs for 
the Lick 2007, Hecto 2010, and MIKE 2016 epochs 
differentially relative to the solar-twin member 
CWW~91 (NID 0739-0790842; EPIC~219698970).
They were observed concurrently (Hectochelle) 
or close in time on the same night,
with the RV zero point of the reference star set to its 
median HARPS RV of $41.654 \pm 0.014$~\kms\ 
(five visits over 1.9~yr). 
For reference, 
\citet{Curtis2013} reported a HIRES epoch of 41.5~\kms\ for 
this reference star.
CWW~91 was not observed on the same run for the MIKE 2012 epoch, 
so instead we calculated the zero point with six other stars with HARPS RVs with 
concurrent MIKE observations
in order to mitigate the effect of any one of those stars being an 
unknown binary. We note that this epoch happens to be the largest outlier, 
although consistent within the estimated uncertainty for our MIKE RVs.

The RVs are provided in Table \ref{t:rvs}. 
Averaging the two Hectochelle RVs, as well as the six HARPS RVs, yields 
six individual RV epochs spanning 9.8 yr with an unweighted rms of 250~\mps.
The HARPS RV rms is 6~\mps\ over 10 months. \\

\begin{deluxetable}{rcccc}
\tabletypesize{\scriptsize}
\tablecaption{RVs for \thisstar \label{t:rvs}}
\tablewidth{0pt}
\tablehead{
\colhead{Date} & \colhead{MJD = JD} & \colhead{RV} & \colhead{Uncertainty} & \colhead{Observatory} \\
\colhead{} & \colhead{$-2,400,000$} & \colhead{(\kms )} & \colhead{(\kms )} & \colhead{}
}
\startdata
2007 Aug 23 & 54335.789 & 41.584 & 1.00  & Lick \\ 
2010 Jul 05 & 55382.264 & 41.397 & 0.30  & Hecto \\ 
2010 Jul 06 & 55383.269 & 41.377 & 0.30  & Hecto \\ 
2012 Sep 30 & 56200.644 & 42.112 & 0.70  & MIKE \\  
2013 Aug 10 & 56514.247 & 41.580 & 0.012 & HARPS \\
2014 May 07 & 56784.386 & 41.586 & 0.008 & HARPS \\
2014 May 08 & 56785.399 & 41.573 & 0.007 & HARPS \\
2014 May 09 & 56786.404 & 41.574 & 0.007 & HARPS \\
2014 May 27 & 56804.311 & 41.570 & 0.016 & HARPS \\
2014 Jun 22 & 56830.298 & 41.577 & 0.008 & HARPS \\ %
2016 Jul 15 & 57584.743 & 41.550 & 0.70  & MIKE \\ 
2017 Jun 02 & 57907.075 & 41.760 & 0.30  & HIRES \\ \hline 
\sidehead{\textit{Star B}:\tablenotemark{a}}
2017 Jun 08 & 57913.062 & $-24.92$ & 0.20 & HIRES \\  
2017 Aug 28 & 57993.804 & $-25.28$ & 0.20 & HIRES \\  
\enddata
\tablecomments{RV measurements collected over nearly ten years, 
with rms = 250~\mps , consistent with \thisstar\ being single. 
See Section \ref{s:sin} for details.} 
\tablenotetext{a}{The faint neighbor referred to as ``Star B'' is
the first object listed in Table~\ref{t:detect} and
located $4''$ south of the exoplanet host at (19:16:22.319, $-$15:46:19.68).}
\end{deluxetable}

\added{
\noindent \textbf{RV median:} 
The median RV of $41.58 \pm 0.25$~\kms\
provides an additional stringent constraint on binarity. 
Consider the Hectochelle RVs: 
of the 50 members observed, 
selecting the 38 stars with RVs within 2~\kms\ of the cluster median, 
the two-night median and standard deviation RV for R147 is $41.384 \pm 0.70$~\kms, 
which is exactly equal to the Hectochelle RV for \thisstar .
Even if this star is single, 
this equality is a coincidence, given R147's intrinsic velocity dispersion. 
The Hectochelle RV spread is likely larger than the intrinsic cluster velocity dispersion due to some binaries lingering in the sample 
and is not yet well-constrained, but 
it is currently estimated to be between $\sigma_{\rm R147} = $0.25-0.50~\kms\ \citep[see Section 3.1.2 in][]{Curtis2016PhD}.

Assuming $M_2 = 0.2$~\msun, RV$_\gamma = $RV$_{\rm R147}$, and $\sigma_{\rm R147} = 0.5~\kms$, 
a hypothetical circular binary seen edge-on would 
require an orbital period $P_{\rm orb} = 1175$~years ($\sim$118~AU)
for the RV semi-amplitude ($K_1$) to match the velocity dispersion. 
Such binaries are ruled out by the AO imaging and coronagraphy, 
except for phases where the projected separation is reduced
under the detection sensitivity curve (dark blue curve in Figure~\ref{f:binary}).
For shorter-period binaries, the RV of the primary will cross the 
cluster's velocity at the conjunction points 
but will be larger or smaller than this value 
during most of the orbital period, neglecting dispersion.
The fact that the RV for \thisstar\ is exactly equal to the 
cluster median 
means that if it is a binary, 
we would be lucky to catch it at conjunction. 

For example, consider once again the hypothetical binary described previously: 
$M_2 = 0.2$~\msun, $e = 0.0$, $i = 90$\degree . 
If the semi-major axis is $a = 45$~AU 
(the approximate boundary of the AO sensitivity curve), 
then $P_{\rm orb}$ = 146~years and $K_1 = 1$~\kms. 
The primary only spends 0.64\% of its orbit within 
the $\sim$10~\mps\ uncertainty of the HARPS RV data.
However, the HARPS RV precision is not the appropriate limit 
because we do not know the intrinsic RV (or center-of-mass velocity, 
RV$_\gamma$, if a binary)
for this star.
If RV$_\gamma \neq \langle$RV$_{\rm obs}\rangle$, 
but instead is some other value within the cluster velocity dispersion, 
then it is possible that we are observing it at a 
quadrature point instead of conjunction, 
which would modestly increase the probability of randomly catching it at this orbital phase 
due to the longer time the star spends at the quadrature RV within the HARPS uncertainty.\\

\noindent \textbf{RV binary constraints:} 
These RVs, particularly the precise measurements from HARPS, 
are useful for constraining binary scenarios with semi-major axes 
closer to the primary than the region probed by AO.
We estimated our detection sensitivity by 
generating simulated RV curves with 
\texttt{RVLIN} \citep{RVLIN}
for binaries with semimajor axes $a < 50$~AU
and secondary masses $M_2 < 0.4$~\msun\ 
(rounding up the 0.34~\msun\ limit derived from photometric modeling).
We performed a simple experiment with circular orbits 
seen edge-on to sketch out the approximate limits on binarity in this region.
For each $M_2$--$a$ combination tested, 
we calculated the orbital period ($P_{\rm orb}$) and 
the primary's velocity semi-amplitude ($K_1$), then 
computed the RV time series with \texttt{RVLIN}. 
Next, we derived the optimal time of periastron passage that 
best aligns the observed RVs to the model, 
which presents a best-case scenario to compute $\chi^2$. 
We decided that a binary was detectable if $\chi_{\rm binary}^2 \geq 2 \, \chi_{\rm single}^2$, 
where the single-star model is a flat line running through the median RV.

The constraints derived from this simple experiment 
are illustrated by green shading in Figure~\ref{f:binary}.
Circular, edge-on binaries with center-of-mass RVs equal to 
the observed median, 
RV$_\gamma = \langle$RV$_{\rm obs}\rangle = 41.58$~\kms,
can be ruled out for most of the remaining parameter space.

Different orbital geometries and 
viewing perspectives will alter the detection sensitivity. 
Eccentricity can increase or decrease our sensitivity depending on 
the specific orbital properties and the phase of the observed RVs.
Inclination decreases sensitivity by reducing the RV semi-amplitude; 
however, it is improbable that the sensitivity would drop to zero, 
because it is unlikely that the binary orbital plane is exactly perpendicular 
to the primary--planet plane.

For now, we will conclude this discussion by stating that 
the evidence suggests that \thisstar\ is likely single.
Further progress can be made by simulating realistic binary systems 
in the cluster and testing them against the observational constraints, 
which is not necessary for this study. 
We already demonstrated that the allowed binary systems would dilute 
the observed transits by a negligible amount. 
As for which component of the hypothetical binary hosts the transits, 
this is accounted for when statistically validating the planet with 
\texttt{BLENDER}, discussed later in Section~\ref{s:blender}, 
by confronting the light-curve with simulations of eclipsing binaries or 
larger planets transiting fainter stars
that are physically associated, or in the background,
to rule out these scenarios.}\\

\added{\subsection{Activity and Rotation}\label{s:hk}
We measured chromospheric \caiihk\ emission indices,
$S$ and \lrphk, 
from our MIKE and Hectochelle spectra
following procedures 
described in \citet{Noyes1984} and \citet{wright2004},
and found $S = 0.208 \pm 0.005$ and 
$\lrphk = -4.80 \pm 0.03$. 
Figure~\ref{f:CaK} shows the Hectochelle \caii~K spectrum 
for \thisstar, 
along with solar spectra 
taken between 2006 and the present, 
which are shaded red to represent the range of the contemporary solar cycle.
The solar spectra were obtained 
by the National Solar Observatory's 
Synoptic Optical Long-term Investigations of the Sun (SOLIS) facility
with the Integrated Sunlight Spectrometer (ISS) on Kitt Peak 
\citep{solis}.\footnote{\url{http://solis.nso.edu/iss}} 
The observed chromospheric activity level of \thisstar\ 
is somewhat higher than the modern solar maximum
\citep[the average maximum over cycles 15--24 is $\lrphk = -4.905$~dex;][]{Egeland2017}, 
which is expected because it is $\sim$1.5 Gyr younger 
than the Sun. 
Applying the activity--rotation--age relation from \citet{mamajek2008}, 
a value of $\lrphk = -4.80$ corresponds to an age of 3.2~Gyr.

\begin{figure}\begin{center}
\includegraphics[width=3.75in,trim=2cm 0 0 0]{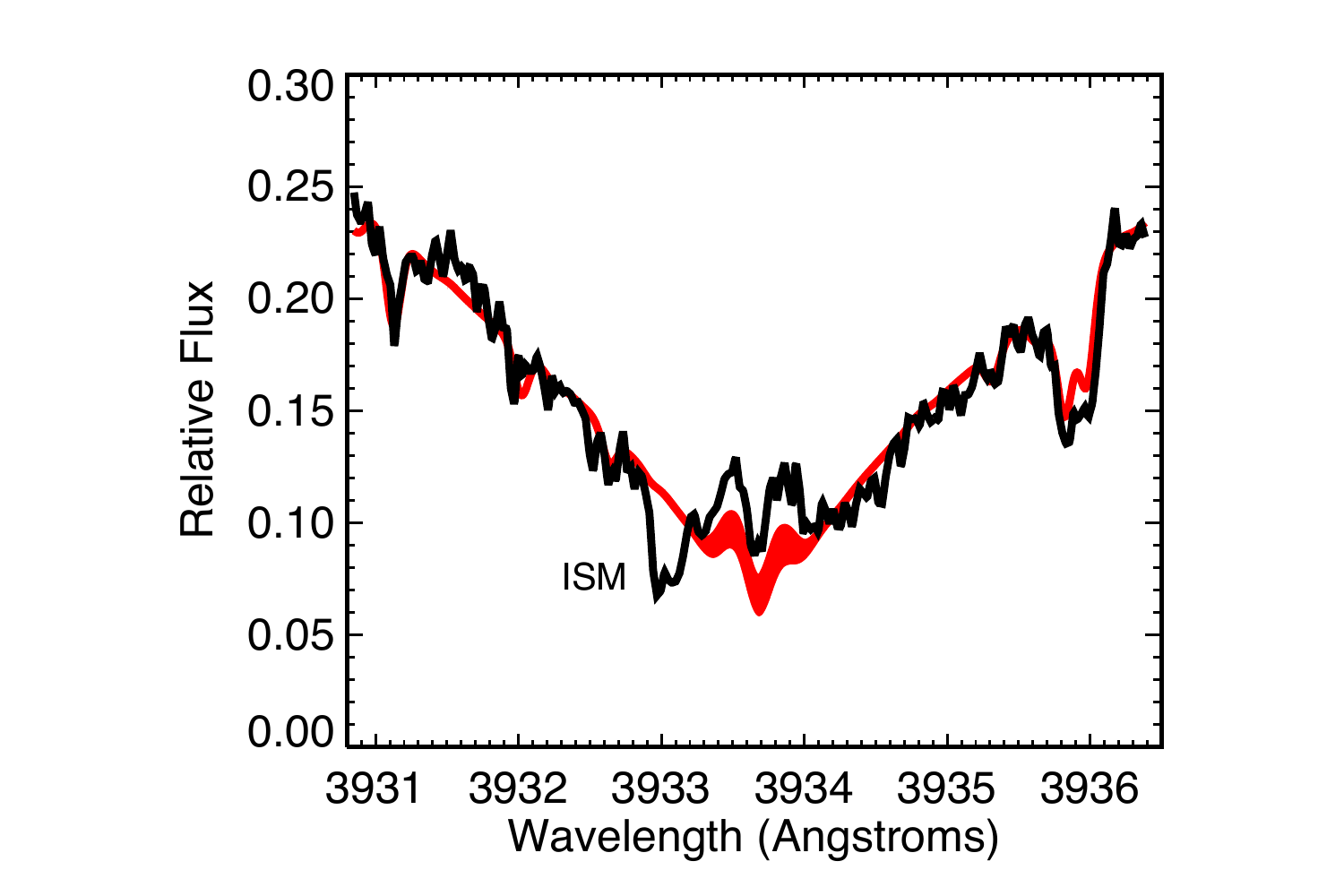}	
	\caption{The \caii~K spectral region for \thisstar\ 
	as observed with MMT/Hectochelle in 2010 July (black line) 
	and SOLIS/ISS spectra of the Sun taken between 
	2006 and the present (red shading)
	to represent the range of the modern solar cycle.
    The chromospheric activity for the 3 Gyr R147 star 
    is slightly elevated above the modern solar maximum, 
	as is typical for this cluster and expected from its age. 
	Note the interstellar absorption line blueward of the 
	\caii~K line core \citep[for more on interstellar 
	absorption and its impact on activity indices, see][]{Curtis2017}.
	\label{f:CaK}}
\end{center}\end{figure}

An analysis of the \caiihk\ activity for the full cluster sample is
underway, 
and these numbers can be considered preliminary until 
that study is complete. 
However, the solar twin status of this star simplifies the 
calibration, as we can tie it directly to 
solar observations. 
We tested this by differentially measuring $S$ for \thisstar\ 
relative to the the SOLIS/ISS spectra 
and applying the conversion from
their 1~\AA\ K-index to $S$ using the 
\citet{Egeland2017} relations. 
This procedure yielded $S = 0.2085$, 
which translates to an approximate increase 
in \lrphk\ over our Hectochelle calibration of only 0.003~dex.
The uncertainties are assessed by considering 
the observed scatter for stars with multiple observations and 
stars with overlapping spectra between MIKE and Hectochelle 
(neglecting astrophysical variability)
and uncertainty in the adopted $(B-V)$ when transforming 
$S$ to $\lrphk$.

The rotation period inferred as part of 
the activity--rotation--age procedure 
\citep[i.e., from the activity--Rossby relation 
combined with the convective turnover time;][]{mamajek2008, Noyes1984}
is $\prot = 21.4$~days.
While spot modulation is clearly evident in the light-curve 
shown in the top panel of Figure~\ref{f:lc}, 
a $\sim$21~day signal is not immediately obvious.
The apparent periodicity is closer to 6--7~days; 
this cannot be the true rotation period because 
the star would correspondingly be much more active, 
with $\lrphk \sim -4.41$ \citep{mamajek2008}.
If there were two major spot complexes on 
opposite sides of the primary star, that would make the period of the modulation be half of the rotation period. 
If the rotational period was actually 12--14 days, 
we would expect $\lrphk = -4.55 \pm 0.05$~dex,\footnote{The conversion from rotation period 
to \lrphk\ depends on the the rotation period and also the adopted $(B-V)$.
The dereddened APASS value is $(B-V)_0 = 0.72$; 
applying the adopted effective temperature to the table of stellar data from 
\citet{Pecaut2013} 
yields $(B-V) = 0.67$.
The uncertainty in each input parameter contributes a similar level of uncertainty.}
which is still too active compared to the observed chromospheric emission.

The \texttt{EVEREST} light-curve was produced with a stationary aperture
that encompassed many bright, neighboring stars. 
However, those same rotation signatures are present
in our $9''$ moving aperture light-curve 
(not shown, but the reader can verify this with the 
light-curve provided),
which means the modulation could instead be attributed to
one of the neighbors in that aperture listed in Table~\ref{t:detect}.
We therefore do not report a rotation period at this time.
This illustrates one of the main challenges
to measuring accurate rotation periods in middle-aged
clusters in crowded fields.}

\section{Planet Validation} \label{s:false}

\begin{figure}\begin{center}
\includegraphics[width=3.5in,trim=1cm 0 0 0]{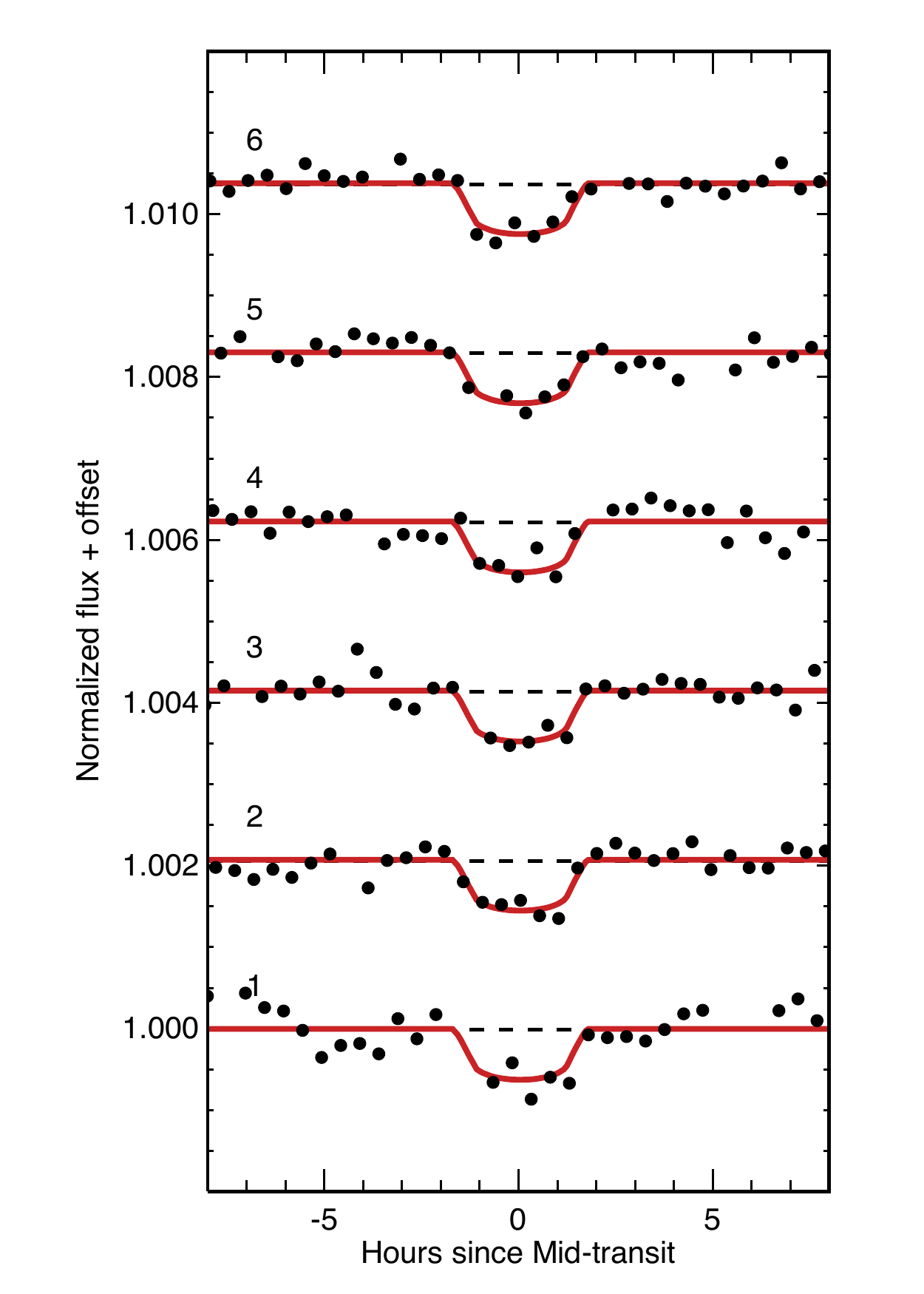}
	\caption{Individual transit events along with the 
	\texttt{EXOFAST} transit model for \thisplanet . 
	Their consistency, especially between odd- and even-numbered events, 
    indicates that they are due to either the same transiting object 
    or two with negligible differences in a circular orbit (i.e., an equal-mass EB).
	\label{f:oddeven}}
\end{center}\end{figure}

First, we inspected the six individual transits for 
variations in depth, timing, and duration between the odd and even events 
that would indicate eccentricity or dissimilar stellar companions,
under the assumption that these are stellar eclipsing binary (EB) transits.
Figure \ref{f:oddeven} shows each transit event separately, 
along with the \texttt{EXOFAST} transit model, and they are all consistent with the model and each other.

One might think that the cluster environment would create a crowded field 
that would complicate the photometric analysis. 
In fact, R147 is relatively sparse due to both  
the low number of (confirmed) members ($N \approx 150$) 
and closer distance compared to clusters like NGC 6811 (295~pc versus $\sim$1100~pc). 
However, R147's location in the Galactic plane near Sagittarius ($l = 21\degree , b = -13\degree$)
means that there are quite a few background stars.
We opted for a circular moving aperture to track 
\thisstar's motion across its individual aperture 
while excluding as many of the background stars shown in Figure~\ref{f:mega} as possible. 
The aperture used to produce the \texttt{EVEREST} light-curve 
that we used to identify the transiting planet 
contained all the bright stars shown to the southwest of \thisstar . 
Our 9$''$ circular aperture excludes all but one of these brighter stars.
We also created apertures as small as 5\farcs5 (1.39 pixels) to reject many of the fainter stars, 
and the transit depth appears the same as in the larger apertures, 
meaning that we can attribute the transit to either 
of the two stars encircled by the dashed line in the figure.


\subsection{Star B: The Bright Neighbor}

The star that remains blended is located approximately 4\farcs2
south of \thisstar , and
we refer to this star as ``star B.''
The mean difference in the various photometric bands shows it to be 3.98~mag fainter than \thisstar\ 
(neglecting differences in interstellar reddening).
The \textit{Gaia} and CFHT/MegaCam epochs are separated by $\sim$6.5~years, 
which is enough to calculate proper motions to test for association with R147, 
given the cluster's relatively large proper motion in declination of $\mu_\delta = -28$~\mas .
For \thisstar , we measure $\mu_\delta = -25.6$~\mas , 
and for star B, we find $\mu_\delta = -9.4$~\mas , 
which does not support cluster membership. 

We can also model the CFHT and UKIRT photometry with 
\texttt{isochrones} under the assumption that it is a single dwarf star 
by applying a Gaussian prior on $\logg = 4.4 \pm 0.5$, 
and we find a 
mass $M = 1.06 -0.10 +0.13$~\msun, 
radius $R = 1.013 -0.16 +0.19$~\rsun, 
distance $d = 2204 -343 +406$~pc, and 
visual extinction $A_V = 0.63 \pm 0.18$~mag. 

The 3D Galactic dust map produced from 2MASS and Pan-STARRS 1 \citep{3Ddust}\footnote{\url{http://argonaut.skymaps.info/query}}
toward \thisstar\ quotes 
an interstellar reddening at 300~pc (the approximate distance to R147) 
of $E(B-V) = 0.07 +0.03 -0.04$ (i.e., $A_V = 0.22 +0.09 -0.12$, 
which is consistent with the value we find from CMD isochrone fitting). 
According to this map, interstellar reddening is $E(B-V) = 0.16 \pm 0.02$ or $A_V = 0.50 \pm 0.06$ at 2.2~kpc, the 
distance we infer for star B, 
and reaches a maximum value of $E(B-V) = 0.17 \pm 0.02$ at 2.28~kpc ($A_V = 0.53$).\footnote{Using the 2015 version 
gives color excesses of 0.05 for R147, 0.18 for star B, and a maximum of 0.20 at 2.44~kpc.}
This value is consistent with our result from \texttt{isochrones} due to the large uncertainty, 
which is compounded when considering our assumption of singularity and a dwarf luminosity class.
The \citet{dustmap} dust map value is marginally less at $E(B-V) = 0.146$ or $A_V = 0.45$, 
and the recalibrated map from \citet{dustmap2} quotes $E(B-V) = 0.125$ or $A_V = 0.39$.

The proper motions and stellar properties are inconsistent with membership, 
meaning that star B is likely a background star. 
A quick test with \texttt{BLENDER} \citep[described in the next section;][]{Blender} 
indicates that the broad features of the transit light-curve 
can indeed be fit reasonably well if star B is a background EB. 
Assuming that both the target and star B are solar-mass stars, 
we find a decent fit for a companion to star B of about 0.26~\msun . 
This EB produces a secondary eclipse, but it is very shallow ($\sim$30~ppm) 
and is probably not detectable in the data, given the typical scatter of $\sim$120~ppm.
If we resolved star B, we expect that the undiluted transit due to this hypothetical EB would be $\sim$2.5\% , 
which could be detected from ground-based photometric observations in and out of transit. 
We attempted to conduct such observations with the 
Las Cumbres Observatory, but were unable to acquire the relevant data. 

Assuming a circular orbit, 
the RV semi-amplitude of 
such a hypothetical single-lined EB is 19.8~\kms ,
which is also feasible to test and rule out with a few RV observations.
We acquired two RV epochs of star B with HIRES, 
which were taken 7.37 and 9.93 days from midtransit (propagated forward according to the 
transit ephemeris in Table~\ref{t:prop}), 
near the secondary eclipse and second quadrature points at phases of 0.53 and 0.72, respectively. 
The RVs, listed at the bottom of Table~\ref{t:rvs}, 
are constant to within their 0.2~\kms\ uncertainties.
Furthermore, these HIRES spectra have sufficient quality to rule out 
secondary spectral lines down to 1\% of the brightness of the primary, 
excluding $\pm$10~\kms\ separation \citep{Kolbl2015}.
This rules out the false-positive scenario where star B is a background EB. 

\subsection{False-alarm Probability} \label{s:blender}
Having excluded the only visible neighboring star 
within the aperture as the source of the transit signal, 
we then examined the likelihood of a false positive caused by unseen stars. 
For this, we applied the \blender\ statistical validation technique 
\citep{Torres2004, Torres2011, Torres2015} that has been used previously to validate
candidates from the {\it Kepler\/} mission 
\citep[see,  e.g.,][]{Torres2017, Fressin2012, Borucki2013, Barclay2013, Meibom2013,
  Kipping2014, Kipping2016, Jenkins2015}. 
  For full details of the methodology and additional examples of its application,
we refer the reader to the first three sources above. Briefly,
\blender\ models the light-curve as a blend between the assumed host
star and another object falling within the photometric aperture that
may be an EB or a star transited by a larger planet,
such that the eclipse depths from these sources would be diluted by
the brighter target to the point where they mimic shallow planetary
transits. These contaminants may be in either the background or
foreground of the target or physically associated with it. Fits to
the {\it K2} light-curves of a large number of such simulated blend
models with a broad range of properties allows us to rule many of them
out that result in poor fits, and Monte Carlo simulations conditioned
on constraints from the follow-up observations (high-resolution
spectroscopy, imaging, RVs, color information) yield a
probability of 99.86\% that the candidate is a planet, as opposed to a
false positive of one kind or another. Thus, we consider
\thisplanet\ 
to be statistically validated as a planet.

\section{Discussion}
We have demonstrated that \thisstar\ is a single, solar twin member of the 
3~Gyr open cluster Ruprecht 147 
and that it 
hosts a statistically validated sub-Neptune exoplanet in a 13.84~day orbit. 

\subsection{Expected yield}
\added{This is the only planetary system found (as of this writing) 
out of 126 RV-confirmed members of R147 
that were observed with \textit{K2} during Campaign 7. 
Neglecting the red giants (eight stars), 
blue stragglers (five stars), 
and tight binaries ($10+$ stars), 
we searched $\sim$100 FGK dwarfs.
According to Table 4 in \citet{Fressin2013}, 
the percentage of stars with at least one planet with 
an orbital period under 29 days is 
0.93\% for giant planets (6--22~\rearth), 
0.80\% for large Neptunes (4--6~\rearth), 
10.24\% for small Neptunes (2--4~\rearth),
12.54\% for super Earths (1.25--2~\rearth), and 
9.83\% for Earth-sized planets (0.8--1.25~\rearth).
If we assume a circular orbit, 
the transit probability is defined as 
the ratio of the sum of the planetary and stellar radii
to the semimajor axis, $P_{tr} \equiv (R_p + R_\star)/a \simeq (R_\star/a)$.
For simplicity, we assume that all stars are the size of the Sun (not too unrealistic).
\citet{Fressin2013} quoted the occurrence rates in 11 period ranges: 
we focus on 
0.8--2.0, 2.0--3.2, 3.2--5.9, 5.9--10, 10--17, and 17--29 days;\footnote{In fact, \citet{Fressin2013} 
quoted the occurrence rates for each period range starting at 0.8~days, 
so we subtract the previous bin's value from the one under consideration. 
For example, the occurrence rate for the 17--29~day bin is the value 
for the 0.8--29~day bin minus the value for the 0.8--17~day bin.}
restricting the orbital periods to $<30$~days 
ensures that at least two transits 
will be present in our $\sim$81~day light-curves.
We calculate transit probabilities for the mean period for each period bin 
and convert these periods to semimajor axes ($a \propto P^{2/3}$) 
to find transit probabilities in each period range.
We estimate the exoplanet yield as 
$N_{\rm planet} = N_{\rm star} \times P_{\rm planet} \times P_{\rm transit} \times P_{\rm detect},$
where $N_{\rm planet}$ is the number of stars observed to host planets with periods under 30 days, 
$N_{\rm star}$ is the number of stars surveyed (100 in this case), 
$P_{\rm planet}$ is the percentage of stars with at least one planet from \citet{Fressin2013}, 
$P_{\rm transit}$ is the transit probability assuming the stars are $R_\star = 1\,R_\odot$, 
and 
$P_{\rm detect}$ is our sensitivity to detecting these transiting planets: we assume that 
we can detect any planet larger than the ``Earth'' class with periods under 30 days. 
Based on this calculation, we expect to detect 
0.05 giants,
0.04 large Neptunes,
0.45 small Neptunes, and
0.66 super Earths,
and we would miss 0.57 Earths, as 
we assume that our survey is not sensitive to the Earth-sized planets \citep{Howell2014}.
Basically, in this RV-vetted sample, we expect our survey to yield $\sim$1 planet, 
which we apparently found.}

\added{As \thisplanet\ was serendipitously discovered by eye 
while browsing light-curves in the course of a stellar rotation period search, 
and not by a pipeline designed to flag planetary candidates, 
we cannot rigorously quantify our detection sensitivity at this time
\citep[e.g.,][]{Rizzuto2016}; 
this is especially important for the Earth and super-Earth classes, 
because these smaller planets might not be so obviously identified visually.
Furthermore, the R147 membership census is incomplete. 
For the ``\textit{K2} Survey of Ruprecht 147,'' 
we allocated apertures based on photometric criteria and soft proper-motion cuts to 
strive for completeness 
and ensure any actual member that is eventually identified
and located in the Campaign~7 field will have a \textit{K2} light-curve.
We selected 1176 stars that passed our tests; 
however, some of these targets are certainly interlopers. 
The impending second \textit{Gaia} data release (DR2) 
will clarify the membership status of 
the majority of these stars.
In the meantime, 
we are working on a new membership catalog that will
supersede \citet{Curtis2013} and include 
detailed stellar properties and
multiplicity informed by AO imaging, RV monitoring, and photometric modeling for 
our expanded RV-vetted membership list \citep{Curtis2016PhD}.
Following the completion of the membership census, 
we will be able to apply our stellar properties derived 
from our vast photometric and spectroscopic database 
to the transit probability calculations and 
incorporate all members with light-curves into our 
occurrence analysis.
Therefore, we opt to postpone a more detailed calculation 
of the exoplanet occurrence rate in R147 until 
these two critical ingredients, membership and sensitivity, 
have been adequately addressed.}

\subsection{Comparison to field stars}
\begin{figure}\begin{center}
\includegraphics[width=3.6in, trim=0.5cm 0 0 0]{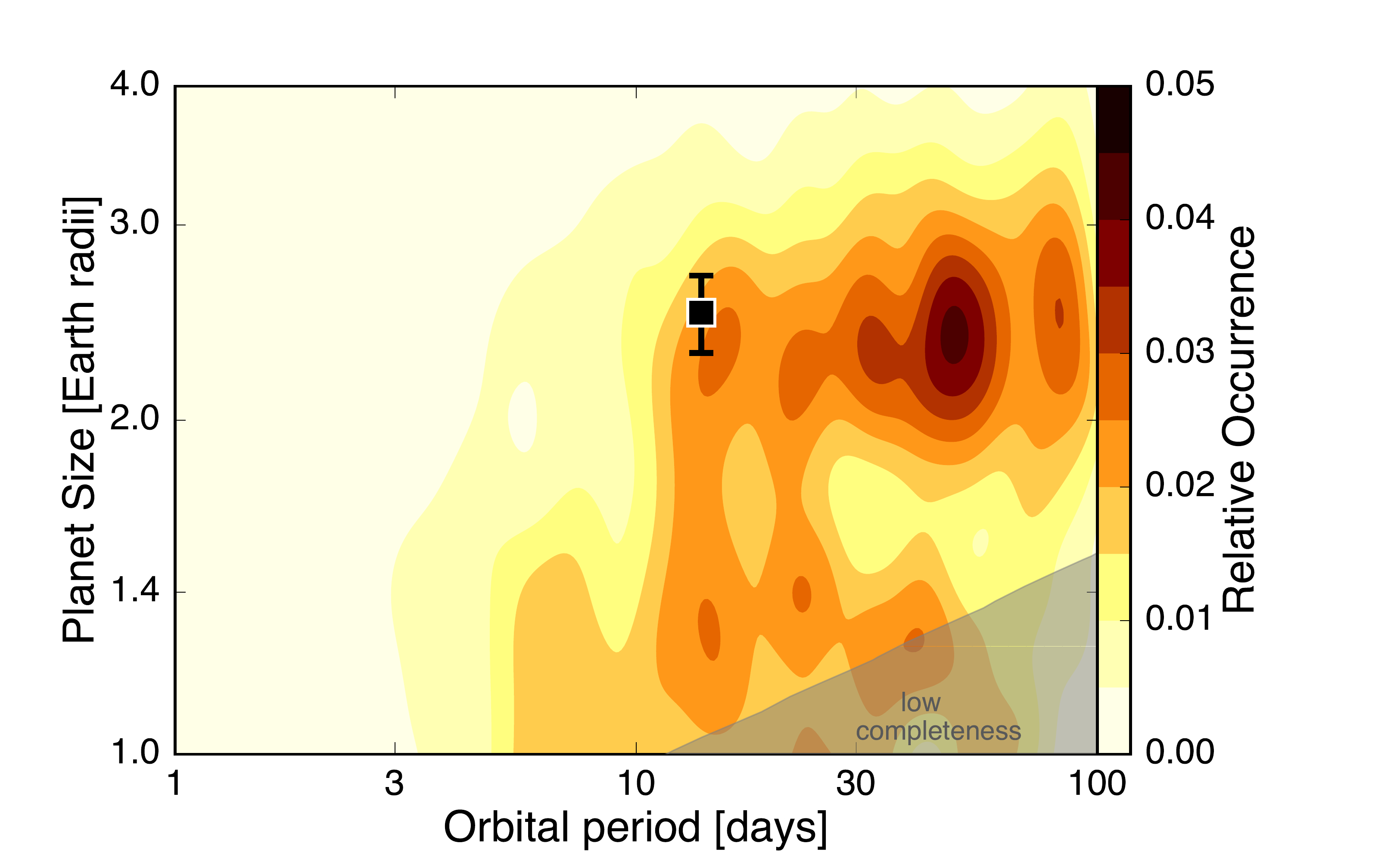}
	\caption{Two-dimensional distribution of planet size and
orbital period found in the \textit{Kepler} field, 
adopted from \citet{Fulton2017}, 
is shown along with the location of \thisplanet. 
Restricting this distribution to periods less than 40 days 
(i.e., demanding the presence of two transits for a planet detection)
means that \thisplanet\ 
is found near a relative maximum in this distribution.
	\label{f:BJ}}
\end{center}\end{figure}

\citet{Fulton2017} showed that the distribution of planetary radii is bimodal, 
with a valley at about 1.8~\rearth\ 
and a peak at the larger side at 2.4~\rearth\ representing sub-Neptunes, 
which they argued are a different class of planets than the super Earths found on 
the smaller side of the gap (see their Figure 7).
With a radius of $\sim$2.5~\rearth, \thisplanet\ falls on the 
large side of the planet radius gap \citep[see also][]{Rogers2015, Weiss2014}.
Our Figure~\ref{f:BJ} presents a modified version of 
the bottom panel of Figure~8 from \citet{Fulton2017}, 
which shows the completeness-corrected, 
two-dimensional distribution of 
planet size and orbital period derived from the 
\textit{Kepler} sample. Our figure compares 
this distribution to the properties of \thisplanet\  
and shows that it is found near a relative maximum.
In other words, \thisplanet\ appears to 
have a fairly typical radius for a short-period ($P<29$~days) planet.

\subsection{Comparison to the NGC~6811 planets}

\citet{Meibom2013} concluded that the 
frequency of planets discovered in the 1~Gyr \textit{Kepler} cluster NGC~6811 
is approximately equal to the \citet{Fressin2013} field rates 
based on two planets found out of 377 members surveyed.
This is about half of the raw rate found in R147 (i.e., 1 in 100 versus 2 in 377); 
in other words, the same order of magnitude.

The two planets found in NGC~6811 are quite similar to \thisplanet: 
they are sub-Neptunes with radii of 2.8 and 2.94~\rearth\ and 
periods of 17.8 and 15.7 days (Kepler-66\,b and 67\,b, respectively). 
This is unlikely to be a mere coincidence, but as Figure~\ref{f:BJ} illustrates, 
planets with these approximate properties are relatively more prevalent. 
However, that figure shows that the relative occurrence of the sub-Neptunes continues, and even increases, 
to longer orbital periods.
While the duration of the \textit{K2} survey of R147 was not long enough to identify planets in the 40--100 day regime, 
presumably such planets could have been found in NGC~6811 during the \textit{Kepler} prime mission. 
With this limited sample, it is unclear if any meaning should be drawn from this 
regarding possible planetary architectures that can form and survive in a dense cluster, 
but it is at the very least an intriguing option to consider.
However, we think this is probably due to the relatively lower $S/N$ light-curves due to NGC~6811's large distance modulus 
and the reduction in transit depth and probability with increasing orbital period.

\subsection{Similar planets and estimating the mass}

Considering the planets with measured masses and radii in the field, 
there are currently five listed on 
\url{exoplanets.org} 
with
$2.4 < R_p / \rearth < 2.7$, 
$K > 1$~\mps , and 
$P > 5$~days: 
\textit{Kepler}'s 96~b, 106~c and e, 131~b, 
and HIP~116454~b.
The basic transit and physical properties of \thisplanet\ and its host are
similar to those of Kepler~106~c: 
$M_\star = 1.0$~\msun, 
[Fe/H]$ = -0.12$~dex,
$\teff = 5860$~K,
$\logg = 4.41$~dex,
$V = 13$,
$P_{orb} = 13$~days, 
$R_P = 2.5$~\rearth, and
$a = 0.111$~AU.
Importantly, the 
RV semi-amplitude for Kepler~106~c is 
$K = 2.71$~\mps, 
and the planet mass is $M_p = 10.4$~\mearth \citep{Marcy2014},\footnote{\url{http://exoplanets.org/detail/Kepler-106_c}} 
and this mass was measured with RV observations made with Keck/HIRES.

Applying the \citet{Wolfgang2016} mass--radius relation for sub-Neptune transiting planets (i.e., $R_P < 4$~\rearth), where 
$M / \mearth = 2.7 (R / \rearth)^{1.3}$, 
predicts a mass for \thisplanet\ of 
$M_p \sim 8.75 \pm 0.9 \pm 1.9$~\mearth , 
where the uncertainties represent the standard deviation of masses computed from a normally distributed 
sample of radii $R_p = 2.5 \pm 0.2$~\rearth\ 
and the normally distributed dispersion in mass of the relation, respectively. 
The \cite{forecaster} probabilistic mass--radius relation, implemented with the \texttt{Forecaster} Python code, 
yields $M_p = 7.2 +5.1 -3.1$~\mearth . 
Assuming a circular orbit, Kepler's Law predicts an RV
semi-amplitude for \thisstar\ of $K \approx 2 \pm 1$~\mps\ in this mass range.
\added{Querying the CPS chromospheric activity catalog 
\citep{Isaacson2010} for dwarfs
with similar color and activity
(i.e., $0.65 < (B-V) < 0.72$, $-4.83 < \lrphk < -4.77$, 
and height above the main sequence $\delta M_V < 1$~mag), 
returns 15 stars with 
measured RV jitters ranging between 2.6 and 3.6~\mps . 
This might be measurable with existing precise RV instruments like HIRES or HARPS, 
as we know the orbit ephemeris and can 
strategically plan repeated observations at quadrature points 
to mitigate the expected jitter.}
\thisplanet\ would then become the first planet with a measured mass and density in an open cluster.

\appendix
\section{Planets discovered in open clusters}

Table~\ref{t:cp} lists the 23 planets and three candidates that have been discovered to date in open clusters. 
We list KIC or EPIC IDs when available, 
whether the planet was discovered via transit or RV techniques (no cluster exoplanet has yet been characterized 
with both techniques), 
the $V$ magnitude and type of host, 
the orbital period, 
the planetary radius or mass ($m\,\sin\,i$), 
citations, and additional notes (e.g., ``HJ,'' referring to hot Jupiter).
We assembled this list to determine how many planets are currently known in clusters, then 
decided that it might be of use and interest to the reader, so we provide it here. 
\added{After we submitted this manuscript, 
\citet{David2018} presented a list 
of ``known and proposed exoplanets in sub-Gyr populations detected via the transit or radial velocity method.'' 
Their Table~1 overlaps considerably with our table 
due to the known cluster planets mostly 
being found in Hyades and Praesepe. 
By construction, their list does not include 
the NGC~6811 or M67 planets (and the R147 planet, since we are announcing it now), 
and we do not list planets found in young associations. 
}
\\

\setcounter{table}{0}
\renewcommand{\thetable}{A\arabic{table}}

\begin{deluxetable*}{lccDcccll}
\tablecaption{Planets in Clusters} \label{t:cp}
\tablehead{\colhead{Planet} & \colhead{KIC/EPIC} & \colhead{Discovery}
& \multicolumn2c{$V$} & \colhead{Period} & \colhead{Radius /} & \colhead{Host}
& \colhead{Notes}
& \colhead{Citations} \\
\colhead{ID} & \colhead{ID} & \colhead{Method}
& \multicolumn2c{(mag)} & \colhead{(days)} & \colhead{$M \sin i$} & \colhead{Info.}
& \colhead{}
& \colhead{}
}
\decimals
\startdata
\sidehead{\textit{Pleiades (130 Myr):}}
$\cdots$ & C4 & $\cdots$ & $\cdots$ & $\cdots$ & $\cdots$ & $\cdots$ & None found & 6 \\ 
\hline\sidehead{\textit{Hyades (650 Myr):}}
$\epsilon$~Tau\,b & 210754593 & RV &  3.53 & 594.9 & 7.6~\mjup   & 2.7~\msun\ Giant & 1st ever & 19 \\ 
HD~285507\,b & 210495452 & RV & 10.47 &  6.09 & 0.917~\mjup & K4.5 & Eccentric HJ & 18 \\ 
K2-25\,b   & 210490365 & Tr & 15.88 & 3.485 & 3.43~\rearth & M4.5 & $\cdots$  & 5, 10 \\ 
K2-136-A\,b & 247589423 & Tr & 11.20 &  7.98 & 0.99~\rearth & K5.5 & Stellar binary & 4, 12 \\ 
K2-136-A\,c & 247589423 & Tr & 11.20 & 17.31 & 2.91~\rearth & K5.5  & Stellar binary & 4, 12 \\ 
K2-136-A\,d & 247589423 & Tr & 11.20 & 25.58 & 1.45~\rearth & K5.5  & Stellar binary & 4, 12 \\ 
HD 283869\,b & 248045685 & Tr & 10.60 & $\sim$106 & 1.96~\rearth & K5 & Candidate (1 transit) & 20 \\
\hline\sidehead{\textit{Praesepe (650 Myr):}}
Pr0201\,b & 211998346 & RV & 10.52 & 4.43    & 0.54~\mjup  & late-F & HJ, ``two b's'' & 17 \\ 
Pr0211\,b & 211936827 & RV & 12.15 & 2.15    & 1.844~\mjup & late-G & HJ, ``two b's'' & 17 \\ 
Pr0211\,c & 211936827 & RV & 12.15 & $>$3500 & 7.9~\mjup   & late-G & Eccentric; 1st multi & 9 \\ 
K2-95\,b  & 211916756 & Tr & 17.27  & 10.14 & 3.7~\rearth & 0.43~\msun & $\cdots$             & 7, 11, 14, 15 \\ 
K2-100\,b & 211990866 & Tr & 10.373 & 1.67  & 3.5~\rearth & 1.18~\msun & $\cdots$             & 1, 7, 11, 16 \\  
K2-101\,b & 211913977 & Tr & 12.552 & 14.68 & 2.0~\rearth & 0.80~\msun & $\cdots$             & 1, 7, 11, 16 \\ 
K2-102\,b & 211970147 & Tr & 12.758 & 9.92  & 1.3~\rearth & 0.77~\msun & $\cdots$             & 11 \\ 
K2-103\,b & 211822797 & Tr & 14.661 & 21.17 & 2.2~\rearth & 0.61~\msun & $\cdots$             & 11 \\ 
K2-104\,b & 211969807 & Tr & 15.770 & 1.97  & 1.9~\rearth & 0.51~\msun & $\cdots$             & 7, 11 \\ 
EPIC~211901114\,b & 211901114 & Tr & 16.485 & 1.65  & 9.6~\rearth & 0.46~\msun & Candidate & 11 \\ 
\hline\sidehead{\textit{NGC 2423 (740 Myr)}\tablenotemark{a}:}
TYC 5409-2156-1\,b & $\cdots$ & RV & 9.45 & 714.3 & 10.6~\mjup & Giant & $\cdots$ & 8 \\ 
\hline\sidehead{\textit{NGC 6811 (1 Gyr):}}
Kepler-66\,b & 9836149 & Tr & 15.3 & 17.82 & 2.80~\rearth & 1.04~\msun & $\cdots$ & 13 \\ 
Kepler-67\,b & 9532052 & Tr & 16.4 & 15.73 & 2.94~\rearth & 0.87~\msun & $\cdots$ & 13 \\ 
\hline\sidehead{\textit{Ruprecht 147 (3 Gyr):}}
\thisplanet & 219800881 & Tr & 12.71 & 13.84 & 2.5~\rearth & Solar twin & $\cdots$ & This work \\ 
\hline\sidehead{\textit{M67 (4 Gyr)\tablenotemark{b}:}}
YBP~401\,b  & $\cdots$       & RV & 13.70 & 4.087 & 0.42~\mjup & F9V   & HJ  & 2, 3 \\ 
YBP~1194\,b & 211411531 & RV & 14.68 & 6.960 & 0.33~\mjup & G5V   & HJ  & 2, 3 \\ 
YBP~1514\,b & 211416296 & RV & 14.77 & 5.118 & 0.40~\mjup & G5V   & HJ  & 2, 3 \\ 
SAND~364\,b & 211403356 & RV &  9.80 & 121   & 1.57~\mjup & K3III & $\cdots$  & 2, 3 \\ 
SAND~978\,b\tablenotemark{c} & $\cdots$       & RV &  9.71 & 511   & 2.18~\mjup & K4III & Candidate  & 2, 3 \\ 
\enddata
\tablerefs{ 
(1) \citet{Barros2016};
(2) \citet{Brucalassi2014};
(3) \citet{Brucalassi2017};
(4) \citet{Ciardi2017}; 
(5) \citet{David2016};
(6) \citet{Gaidos2017};
(7)  \citet{Libralato2016};
(8) \citet{LovisMayor2007};
(9) \citet{Malavolta2016};
(10) \citet{Mann2016};
(11) \citet{Mann2017};
(12) \citet{Mann2017Hyades};
(13) \citet{Meibom2013};
(14) \citet{Obermeier2016};
(15) \citet{Pepper2017};
(16) \citet{Pope2016};
(17) \citet{Quinn2012};
(18) \citet{Quinn2014};
(19) \citet{Sato2007};
(20) \citet{Vanderburg2018Hyad}.
}
\tablenotetext{a}{\citet{LovisMayor2007} also announced a substellar object in NGC 4349,
but it has a minimum mass of 19.8~\mjup, greater than the planet--brown dwarf boundary at 11.4--14.4~\mjup, and so we do not include it here.}
\tablenotetext{b}{\citet{Nardiello2016} announced some candidates, which they concluded are likely not members of M67.}
\tablenotetext{c}{\citet{Brucalassi2017} referred to this detection as a planet candidate and 
stated that YBP~778 and YBP~2018 are also promising candidates.}
\label{t:cp}
\end{deluxetable*}

\acknowledgments

A.~Vanderburg produced the light-curve used in this work and corroborated the initial discovery; 
he is supported by the NASA Sagan Fellowship.
G.~Torres performed the \texttt{BLENDER} false-alarm analysis, 
and acknowledges partial support for this work from NASA grant
NNX14AB83G ({\it Kepler\/} Participating Scientist Program).
A.W.~Howard led the acquisition of HIRES spectra of the 
planet host and the faint neighbor.
H.~Isaacson measured stellar RVs for those targets and 
checked the spectra for secondary light. 
D.~Huber provided the access to Keck/HIRES needed to acquire those spectra  
and also performed the \texttt{isoclassify} analysis; 
he acknowledges support by the National Aeronautics and Space Administration under grant NNX14AB92G 
issued through the \textit{Kepler} Participating Scientist Program.
A.L.~Kraus, A.C.~Rizzuto, and A.W.~Mann
acquired and analyzed the Keck adaptive optics data.
A.W.~Mann fit the light-curve to measure the transit properties and contributed Figure~\ref{f:corner}. 
A.L.~Kraus obtained the UKIRT/WFCAM imaging.
B.J.~Fulton contributed Figure~\ref{f:BJ}.
C.~Henze ran the \texttt{BLENDER} jobs on the Pleiades supercomputer 
and pre-processed the output.
J.T.~Wright advised the Ph.D. dissertation work of J.L.~Curtis, 
which amassed much of the basic data presented herein 
(e.g., proprietary photometry and spectroscopy), and 
is the submitting and administrative PI of the \textit{K2} program GO 7035. 

The remainder of the work was completed by J.L.~Curtis, 
including the planet discovery, 
host star characterization,
preliminary transit fitting, 
and synthesis of the data and contributions provided by the coauthors.
He successfully led a petition for Campaign 7 to point at Ruprecht 147 
and, as science PI of GO 7035, was awarded the program to survey the cluster while 
a graduate student at Penn State University and 
member of the Center for Exoplanets and Habitable Worlds.
He was granted access to Magellan while serving as an 
SAO predoctoral fellow at the Harvard--Smithsonian Center for Astrophysics. 
The planet discovery and characterization work was performed after 
joining Columbia University.

J.L.~Curtis is supported by the National Science Foundation 
Astronomy and Astrophysics Postdoctoral Fellowship under award AST-1602662
and
the National Aeronautics and Space Administration under
grant NNX16AE64G issued through the \textit{K2} Guest Observer Program (GO 7035).
He thanks the referee for their feedback,
Jason T. Wright and Marcel Ag\"{u}eros for serving as his mentors, 
Luca Malavolta for commenting on a draft of this manuscript and
for providing early access to the HARPS RVs acquired by the Minniti team, 
Iv\'{a}n Ram\'{i}rez and Luca Casagrande for providing their temperature measurements, 
Fabienne Bastien and Jacob Luhn for commenting on a draft of this manuscript and 
discussing RV jitter,
the Harvard--Smithsonian Center for Astrophysics 
telescope allocation committee for granting access to Magellan,
the \textit{K2} Guest Observer office and Ball Aerospace for 
re-positioning the Campaign 7 field to accommodate Ruprecht 147,
and the coinvestigators of the ``K2 Survey of Ruprecht 147'' (GO 7035):
Jason T. Wright, Fabienne Bastien, \soren , Victor Silva Aguirre, and Steve Saar. 

The Center for Exoplanets and Habitable Worlds is supported by the
Pennsylvania State University, the Eberly College of Science, and the
Pennsylvania Space Grant Consortium.


This paper includes data collected by the \textit{K2} mission. 
Funding for the \textit{Kepler} and \textit{K2} missions is provided by the NASA Science Mission directorate.
We obtained these data from the Mikulski Archive for Space Telescopes (MAST). 
STScI is operated by the Association of Universities for Research in Astronomy, Inc., 
under NASA contract NAS5-26555. 
Support for MAST for non-HST data is provided by the NASA Office of Space Science via 
grant NNX09AF08G and by other grants and contracts.

This research has made use of the Periodogram Service of the 
NASA Exoplanet Archive, which is operated by the California Institute of Technology, under contract with the National Aeronautics and Space Administration under the Exoplanet Exploration Program.

This research has made use of the Exoplanet Orbit Database
and the Exoplanet Data Explorer at \url{exoplanets.org} \citep{exoplanetsorg}.

This work is based on observations obtained with MegaCam, 
a joint project of CFHT and CEA/DAPNIA, at the
Canada--France--Hawai`i Telescope (CFHT), which is operated
by the National Research Council (NRC) of Canada, the Institute
National des Sciences de l'Univers of the Centre National
de la Recherche Scientifique of France, and the University of
Hawai`i. Observing time was granted by the University of Hawai`i
Institute for Astronomy TAC. These data were reduced at the
TERAPIX data center located at the Institut d'Astrophysique de
Paris.

This publication makes use of data products from the Two
Micron All Sky Survey, which is a joint project of the University
of Massachusetts and the Infrared Processing and Analysis
Center/California Institute of Technology, funded by NASA
and the NSF.

This research was made possible through the use of the AAVSO Photometric All-Sky Survey (APASS), funded by the Robert Martin Ayers Sciences Fund.

This publication makes use of data acquired from UKIRT while it was operated by the Joint Astronomy Centre on behalf of the Science and Technology Facilities Council of the U.K.
UKIRT is supported by NASA and operated under an agreement among the University of Hawai`i, the University of Arizona, and Lockheed Martin Advanced Technology Center; operations are enabled through the cooperation of the East Asian Observatory.

This work has made use of data from the European Space Agency (ESA)
mission {\it Gaia},
processed by
the {\it Gaia} Data Processing and Analysis Consortium 
(DPAC).\footnote{\url{https://www.cosmos.esa.int/web/gaia/dpac/consortium}}
Funding
for the DPAC has been provided by national institutions, in particular
the institutions participating in the {\it Gaia} Multilateral Agreement.

This publication makes use of data products from the Wide-field Infrared Survey Explorer, which is operated by the Jet Propulsion Laboratory, California Institute of Technology, under contract with or with funding from the National Aeronautics and Space Administration. This research has also made use of NASA's Astrophysics Data System, 
and the VizieR and SIMBAD databases, operated at CDS, Strasbourg, France.

This research has made use of the ESO Science Archive Facility to access 
data collected for ESO programmes 091.C-0471(A) and 095.C-0947(A).

Some of the data presented herein were obtained at the W.M. Keck Observatory, which is operated as a scientific partnership among the California Institute of Technology, the University of California, and the National Aeronautics and Space Administration. The observatory was made possible by the generous financial support of the W.M. Keck Foundation.
We wish to recognize and acknowledge the very significant cultural role and reverence that the summit of Maunakea has always had within the indigenous Hawai`ian community.  We are most fortunate to have the opportunity to conduct observations from this mountain.

This work also utilized SOLIS data obtained by the NSO Integrated Synoptic Program (NISP), 
managed by the National Solar Observatory, 
which is operated by the Association of Universities for Research in Astronomy (AURA), Inc., 
under a cooperative agreement with the National Science Foundation.

\vspace{5mm}
\facilities{Kepler (K2), 
            CFHT (MegaCam), 
            ESO:3.6m (HARPS),
            Keck:I (HIRES),
            Keck:II (NIRC2),
            Magellan:Clay (MIKE),
            MMT (Hectochelle),
            Shane (Hamilton),
            SOLIS (ISS),
            UKIRT (WFCAM)}


\software{BARYCORR \citep{barycenter}, 
        batman \citep{batman},
        EXOFAST \citep{EXOFAST},  
        forecaster \citep{forecaster},
        isochrones \citep{Morton2015},
        isoclassify \citep{isoclassify},
        RVLIN \citep{RVLIN},
        SME \citep{sme}
             }


\bibliographystyle{aasjournal}

\end{document}